%% file: master.tex
\documentclass[10pt,journal,compsoc]{IEEEtran}
\usepackage[flushleft]{threeparttable}
\usepackage{lipsum,afterpage,refcount}
\usepackage{tablefootnote}
\usepackage{amsmath,amssymb}
\usepackage{siunitx}
\usepackage[T1]{fontenc}
\usepackage{caption}
\usepackage{booktabs}  % nice looking tables
\usepackage{booktabs, multirow} % for borders and merged ranges
\usepackage{changepage,threeparttable} % for wide tables
\usepackage{graphicx}
\usepackage{fancyheadings}
\usepackage{setspace}
\usepackage{color}
\usepackage{url}
\usepackage{hyperref}
\usepackage{amsthm}
\usepackage{amsmath, bm}
\usepackage{mathtools}
\usepackage{flexisym}
\theoremstyle{definition}

\usepackage{multirow}

\usepackage[flushleft]{threeparttable}
\usepackage{color}
\usepackage[table]{xcolor} % for cell colors
\usepackage{array}
\newcolumntype{P}[1]{>{\centering\arraybackslash}p{#1}}
\newcolumntype{M}[1]{>{\centering\arraybackslash}m{#1}}
\usepackage{stfloats}
\usepackage{changepage,threeparttable} % for wide tables
\usepackage{subcaption}
\usepackage{algpseudocode}
\usepackage{amssymb}
\usepackage{enumitem}
\usepackage{booktabs}
\usepackage{verbatim}
\usepackage{fixltx2e}
\usepackage{amsmath}
\usepackage{algorithm}
\usepackage{lipsum}
\usepackage{textcomp}
\usepackage{tikz}
\usepackage{bbding,enumitem}
\usepackage{xspace}
\usepackage{academicons}
\makeatletter
\global\let\tikz@ensure@dollar@catcode=\relax
\makeatother
\DeclarePairedDelimiter{\ceil}{\lceil}{\rceil}
\AtBeginDocument{
    \catcode`_=12
    \begingroup\lccode`~=`_
    \lowercase{\endgroup\let~}\sb
    \mathcode`_="8000
}

\ifCLASSOPTIONcompsoc
\usepackage[nocompress]{cite}
\else
\usepackage{cite}
\fi

\DeclarePairedDelimiterX\set[1]\lbrace\rbrace{#1}

\newcommand{\orcid}[1]{\href{https://orcid.org/#1}{\textcolor[HTML]{A6CE39}{\includegraphics[width=1.7ex]{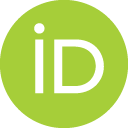}}}}

\newcommand*{\Mem}{Memory Unit\xspace}

\newcolumntype{L}{>{\centering\arraybackslash}p{0.58cm}}
\newcolumntype{Z}{>{\centering\arraybackslash}p{.95cm}}
\newcolumntype{A}{>{\centering\arraybackslash}p{.85cm}}
\newcolumntype{B}{>{\centering\arraybackslash}p{1.15cm}}
\newcolumntype{C}{>{\centering\arraybackslash}p{1.25cm}}
\newcolumntype{D}{>{\centering\arraybackslash}p{1.3cm}}
\newcolumntype{E}{>{\centering\arraybackslash}p{2.7cm}}
\newcolumntype{F}{>{\centering\arraybackslash}p{1.7cm}}
\newcolumntype{H}{>{\setbox0=\hbox\bgroup}c<{\egroup}@{}}

\newcommand*{\threefrac}[3]{%
  \begin{array}{@{\,}c@{\,}}%
    #1\\
    \hline
    #2\\
    \hline
    #3%
  \end{array}%
}

\newcommand{\eq}[1]{Eq. ({#1})\xspace}
\newcommand{\equations}[1]{Eqs. ({#1})\xspace}
\newcommand{\eqPost}[1]{({#1})\xspace}
\newcommand*{\Figure}{Fig.\xspace}
\newcommand*{\Figures}{Figs.\xspace}
\newcommand*{\fig}{\Figure}
\newcommand*{\secn}{Section\xspace}

\newcommand*{\citeSysArr}{\cite{tpu-v1} \cite{tpu-v3} \cite{tpu-v4} \cite{cong-sa} \cite{zhang2019caffeine}\xspace}

\newcommand*{\thpt}{GOPS \tnote{1}}
\newcommand*{\gpmac}{GOPS/ \tnote{2} multiplier }
\newcommand*{\macUt}{Operations/  \tnote{3} multiplier/ clock cycle}

\newcommand*{\thptExpl}{\item[1] Throughput, defined in \eq{\ref{metric:thpt}}, motivation and explanation provided in \secn \ref{sec:metrics}.}
\newcommand*{\macPerDspExpl}{\item[2] Throughput per compute area, defined in
  \eq{\ref{metric:thpt-area}}, motivation and explanation provided in \secn \ref{sec:metrics}.}
\newcommand*{\macUtExpl}{\item[3] Throughput per compute area per clock cycle, defined in \eq{\ref{metric:thpt-area-cc}}, motivation and explanation provided in \secn \ref{sec:metrics}.}
\newcommand*{\footNoteRefs}{\item[1-3] See the corresponding definitions from Table \ref{tab:first}.}

\newcommand*{\multDominant}{\cite{liu2021winocnn} \cite{tpu-v1} \cite{tpu-v3}\xspace}
\newcommand*{\multComplexity}{\cite{fpga_nn_survey} \cite{leon2018walking}  \cite{Pekmestzi1999}\xspace}
\newcommand*{\winoConv}{\cite{lavin_fast} \cite{min_filt}\xspace}

\newcommand*{\spcc}{\text{ }\text{ }}
\newcommand*{\spc}{\text{ }}

\newcommand*{\eOps}          {1.4-1.8\x}
\newcommand*{\eOpsArea}      {3.7-4.6\x}
\newcommand*{\eOpsAreaCC}    {1.6-2\x}

\newcommand*{\sOps}          {1.1-1.4\x}
\newcommand*{\sOpsArea}      {1.4-1.8\x}

\newcommand*{\mixedOps}      {1.7-4.6\x}
\newcommand*{\mixedOpsArea}  {2.8-6\x}
\newcommand*{\mixedOpsAreaCC}{1.6-3.7\x}

\newcommand*{\eALMs}{118K}
\newcommand*{\eRegs}{311K}
\newcommand*{\eMems}{1782}
\newcommand*{\eDSPs}{1072}
\newcommand*{\eFreq}{388}
\newcommand*{\eAlexNetGOPS}{2277}
\newcommand*{\eAlexNetGOPSArea}{1.062}
\newcommand*{\eAlexNetGOPSAreaCC}{2.739}
\newcommand*{\eResNetAGOPS}{2529}
\newcommand*{\eResNetAGOPSArea}{1.180}
\newcommand*{\eResNetAGOPSAreaCC}{3.042}
\newcommand*{\eResNetBGOPS}{2752}
\newcommand*{\eResNetBGOPSArea}{1.284}
\newcommand*{\eResNetBGOPSAreaCC}{3.310}
\newcommand*{\eResNetCGOPS}{2838}
\newcommand*{\eResNetCGOPSArea}{1.324}
\newcommand*{\eResNetCGOPSAreaCC}{3.414}

\newcommand*{\sALMs}{199K}
\newcommand*{\sRegs}{530K}
\newcommand*{\sMems}{2713}
\newcommand*{\sDSPs}{1072}
\newcommand*{\sFreq}{346}
\newcommand*{\sAlexNetGOPS}{1974}
\newcommand*{\sAlexNetGOPSArea}{0.921}
\newcommand*{\sAlexNetGOPSAreaCC}{2.659}
\newcommand*{\sResNetAGOPS}{2258}
\newcommand*{\sResNetAGOPSArea}{1.053}
\newcommand*{\sResNetAGOPSAreaCC}{3.042}
\newcommand*{\sResNetBGOPS}{2458}
\newcommand*{\sResNetBGOPSArea}{1.146}
\newcommand*{\sResNetBGOPSAreaCC}{3.311}
\newcommand*{\sResNetCGOPS}{2534}
\newcommand*{\sResNetCGOPSArea}{1.182}
\newcommand*{\sResNetCGOPSAreaCC}{3.413}

\newcommand*{\ops}{\textit{throughput}\xspace}
\newcommand*{\opsArea}{\textit{throughput per compute area}\xspace}
\newcommand*{\opsAreaCC}{\textit{throughput per compute area per clock cycle}\xspace}

\newcommand*{\opsDSP}{\textit{throughput/DSP}\xspace}
\newcommand*{\opsMult}{\textit{throughput/multiplier}\xspace}

\newcommand*{\x}{$\times$\xspace}
\newcommand*{\by}{$\times$}
\newcommand*{\ex}{\cdot}

\newcommand*{\freq}{f}
\newcommand*{\thpte}{op/s}
\newcommand*{\roof}{roof\xspace}
\newcommand*{\thRoof}{op/s\spc roof}
\newcommand*{\ea}{\textit{et al.}\xspace}
\newcommand*{\gx}{Arria 10 GX 1150\xspace}

\newcommand*{\sx}{Arria 10 SX 660\xspace}
\newcommand*{\tip}{baseline\xspace}

\newcommand*{\TIP}{Baseline\xspace}

\newcommand*{\traditional}{traditional\xspace}

\newcommand*{\lb}{\left(}
\newcommand*{\rb}{\right)}
\newcommand\NoDo{\renewcommand\algorithmicdo{}}

\newcommand\copyrighttext{%
  \scriptsize \textcopyright
  2023 IEEE. Personal use of this material is permitted. Permission
  from IEEE must be obtained for all other uses, in any current or future
  media, including reprinting/republishing this material for advertising or
  promotional purposes, creating new collective works, for resale or
  redistribution to servers or lists, or reuse of any copyrighted
  component of this work in other works.

  Accepted for publication in IEEE Transactions on Computers. DOI: 10.1109/TC.2023.3334140}
\newcommand\copyrightnotice{%
  \begin{tikzpicture}[remember picture,overlay]
    \node[anchor=south,yshift=0pt] at (current page.south) {\fbox{\parbox{\dimexpr\textwidth-\fboxsep-\fboxrule\relax}{\copyrighttext}}};
  \end{tikzpicture}
  \vspace{-11.94pt}
}

\begin{document}

\title{Fast Inner-Product Algorithms and Architectures for Deep Neural Network Accelerators}

\author{Trevor~E.~Pogue~\orcid{0000-0002-6791-3758},~Nicola~Nicolici~\orcid{0000-0001-6345-5908},~\IEEEmembership{Senior Member,~IEEE}
\IEEEcompsocitemizethanks{\IEEEcompsocthanksitem T. E. Pogue and N. Nicolici are with the Department of Electrical and Computer Engineering, McMaster University, Hamilton, ON, L8S 4L8, Canada \protect\\
Email: poguete@mcmaster.ca; nicolici@mcmaster.ca}}

\IEEEpeerreviewmaketitle

\IEEEtitleabstractindextext{%
\begin{abstract}
    \input{abstract}
\end{abstract}
}

\maketitle
\IEEEdisplaynontitleabstractindextext
\IEEEpeerreviewmaketitle
\input{intro}
\input{background-and-related-work}
\input{algorithms}
\input{architectures}
\input{system}
\input{results}
\input{conclusion}

\bibliographystyle{IEEEtran}
\bibliography{bibl}

\begin{IEEEbiography}[{\includegraphics[width=1in,height=1.25in,clip,keepaspectratio]{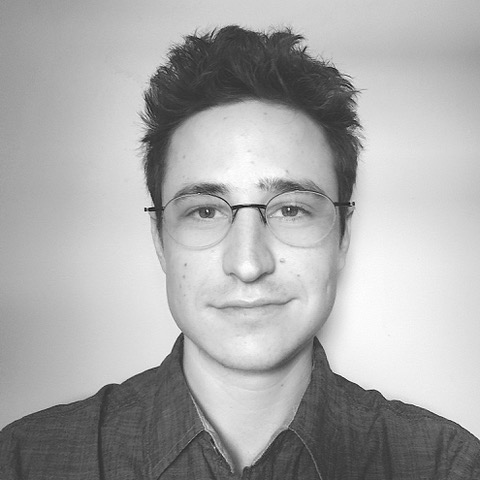}}]{Trevor E. Pogue}
  Trevor E. Pogue received the B.Eng. degree in Electrical Engineering and the
  M.A.Sc. degree in Electrical and Computer Engineering from McMaster
  University, Hamilton, Canada, in 2016 and 2019, respectively. In 2018, he
  worked as a design verification intern at Synopsys Inc., and in 2022-2023
  he worked as an intern at Advanced Micro Devices Inc. He is currently a Ph.D.
  student in the
  Department of Electrical and Computer
  Engineering at McMaster University, Hamilton, Canada.
  His research interests are in the area of hardware acceleration.
\end{IEEEbiography}

\begin{IEEEbiography}[{\includegraphics[width=1in,height=1.25in,clip,keepaspectratio]{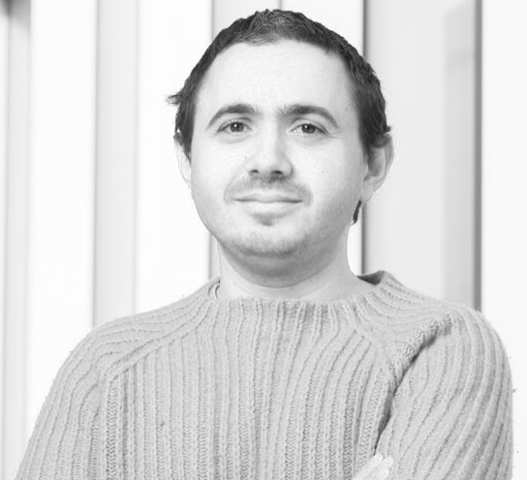}}]{Nicola
    Nicolici}(S’99-M’00-SM'11)
  Nicola Nicolici (S99-M00-SM’11) received the Dipl.Ing. degree in Computer
  Engineering from the “Politehnica” University of Timisoara, Romania, in 1997
  and the Ph.D. degree in Electronics and Computer Science from the University
  of Southampton, U.K., in 2000. He is currently a Professor with the Department
  of Electrical and Computer Engineering, McMaster University, Hamilton, Canada.
  His research interests are in the area of computer-aided design and test. He
  has authored a number of papers in this area. Dr. Nicolici was the recipient
  of the IEEE TTTC Beausang Award for the Best Student Paper at the
  International Test Conference in 2000 and the Best Paper Award at the IEEE/ACM
  Design Automation and Test in Europe Conference in 2004.
\end{IEEEbiography}

\end{document}

%% file: abstract.tex
We introduce a new algorithm called the Free-pipeline Fast Inner Product (FFIP)
and its hardware architecture that improve an under-explored fast inner-product
algorithm (FIP) proposed by Winograd in 1968. Unlike the unrelated Winograd
minimal filtering algorithms for convolutional layers, FIP is applicable to all
machine learning (ML) model layers that can mainly decompose to matrix
multiplication,
including fully-connected, convolutional, recurrent, and attention/transformer
layers. We implement FIP for the first time in an ML accelerator then present
our FFIP algorithm and generalized architecture which inherently improve FIP's
clock frequency and, as a consequence, throughput for a similar hardware cost.
Finally, we contribute ML-specific optimizations for the FIP and FFIP algorithms
and architectures. We show that FFIP can be seamlessly incorporated into
traditional fixed-point systolic array ML accelerators to achieve the same
throughput with half the number of multiply-accumulate (MAC) units, or it can
double
the maximum systolic array size that can fit onto devices with a fixed hardware
budget. Our FFIP implementation for non-sparse ML models with 8 to 16-bit
fixed-point inputs achieves higher throughput and
compute efficiency than the best-in-class prior solutions on the same type of
compute platform.

%% file: intro.tex
\IEEEraisesectionheading{\section{Introduction}\label{sec:intro}}
\copyrightnotice

\IEEEPARstart{R}{ecent} years have seen many works on system-level improvements
for machine learning hardware acceleration and hardware-oriented deep neural
network (DNN) model optimizations. At a certain point, however, after
hardware-oriented DNN model optimizations reach their limit, after the known
parallelism and system-level optimizations for executing their compute patterns
are exploited, and after technology scaling slows to a halt, there is an
accelerator wall which causes limited improvement on the implementation side
\cite{wall}. A less-explored avenue to continue advancement after this point is
to reduce the workload at the algebraic level, by calculating the same machine
learning (ML) model
algebra, nevertheless using a re-arranged compute pattern which produces the
same output from fewer or cheaper operations performed in hardware.

This type of arithmetic complexity research has been explored with application
to deep learning quite extensively in the case of Winograd's minimal filtering
algorithms applied to convolutional neural networks
(CNN)s \winoConv. However, Winograd also introduced an \textbf{unrelated and
less-explored} algorithm in 1968 on which our work is based that reduces the
arithmetic complexity of general matrix multiplication. We refer to this
under-explored algorithm from 1968 as the Fast Inner Product (FIP). Compared to
the \traditional inner product (referred to as \tip), FIP allows matrix
multiplication to be performed with approximately half of the
multiply-accumulate (MAC) operations traded for low-bitwidth additions. Since
MAC units are commonly the area-dominant computational resource in ML
accelerators \multDominant, cutting the number of required MAC units in half
could nearly double the performance per compute area.

Despite the high impact potential of these benefits, the FIP algorithm
\cite{wino} has never before been implemented in a machine learning hardware
accelerator. The first step in our work is to address this oversight and carry
out that investigation. We show that, while the number of required MAC units is
nearly halved as expected, there is also a reduction in clock
frequency and, as a consequence, throughput.

In order to address this weakness of FIP, we introduce a new inner-product
algorithm referred to as the Free-pipeline Fast Inner Product (FFIP) that
achieves the same near 2\x reduction in required MAC units, however, it
inherently improves the clock frequency and, as a consequence, overall
throughput for a similar hardware cost compared to FIP. Additionally, as
expected from our theoretical analysis, the effective size of the largest
systolic array that could be fit in our experimental compute platform was
increased from 56\by56 processing elements (PE)s when using a baseline systolic
array to 80\by80 PEs when using (F)FIP systolic arrays, an increase of over 2\x
in effective number of PEs.

Our work shows that the FFIP architecture is functionally equivalent to
traditional systolic array architectures, and that it can be seamlessly
incorporated into any ML accelerator system
that uses traditional fixed-point systolic arrays for the arithmetic to double
the throughput per MAC unit and significantly increase its performance per
compute area across all ML models that will execute on the systolic array. This
is because any such accelerator could substitute its traditional systolic array
PEs for just nearly half the number of proposed FFIP PEs without fundamentally
altering the accelerator's functionality or internal interfaces in any way.
Furthermore, systolic array accelerators have been proven effective for
accelerating a wide range of modern ML models \citeSysArr.
Our results indicate that FFIP, when overlaid on top of the
most efficient systolic array accelerator systems used in practice,
can further improve compute efficiency and increase the theoretical
performance limits across a wide
range of devices, system implementations, and ML models.

In summary, our key contributions are the following:
\begin{itemize}
\item We implement FIP for the first time in a machine learning accelerator and
validate our generalized analysis of its ability to increase an accelerator's
theoretical throughput and compute efficiency limits for the data types commonly
used in ML acceleration. As part of this undertaking, we also identify a
weakness in FIP of a reduction in clock frequency. To address this, we introduce
a new inner-product algorithm (FFIP) in \secn \ref{sec:wino-improved} to
inherently bypass this trade-off. Additionally, we provide ML-specific
optimizations for both the FIP and FFIP algorithms in \secn
\ref{sec:alg-ml-opt}.
\item We propose a generalized PE architecture in \secn \ref{sec:pe} for
performing FFIP in hardware which inherently improves the clock frequency and,
as a consequence, overall throughput for a similar hardware cost compared to
FIP. Additionally, we provide ML-specific hardware optimizations for the
systolic array architectures housing FIP or FFIP PEs in \secn \ref{sec:ml-mxu}.
We demonstrate in \secn \ref{results} how an example implementation of our
abstract FFIP architecture achieves higher throughput and compute efficiency
across different ML models than the best-in-class prior solutions implemented on
the same type of compute platform.
More importantly, the results indicate that when FFIP is overlaid on top of the
most efficient systolic array systems used in practice, it can further increase
compute efficiency in the general case across a wide range of devices, system
implementations, and ML models.
\end{itemize}

%% file: background-and-related-work.tex
\section{Related Work and Background}
The architectures evaluated and proposed in this work are systolic arrays, which
will also be referred to as matrix multiplication units (MXU)s. Systolic
arrays are an effective choice for use in machine learning accelerators as they
significantly reduce the required memory traffic and can reach higher clock
frequencies than other designs due to the short and regular nature of their
interconnects. This type of architecture has been used in state-of-the-art
deep learning accelerators such as the Tensor Processing Unit (TPU)
\cite{tpu-v1} \cite{tpu-v3} \cite{tpu-v4}, among others \cite{cong-sa}
\cite{zhang2019caffeine}.

\subsection{Related Work}
The FIP algorithm \cite{wino} has been explored by Gustafsson \ea
for application to finite
impulse response (FIR) filtering \cite{wino_fir} in a non-systolic-array
architecture proposal. Three prior works have also been proposed for
systolic array architectures for exploiting FIP 
\cite{wino_sys_1level} \cite{stanford} \cite{mm_fpga_nogain}.

However, none of the prior works on FIP architectures are focused on application
to
machine learning, which uses a certain range of matrix sizes and data
representations, making it deserving of a stand-alone study. For all prior works
related to ML acceleration which cite FIP \cite{wino}, upon closer inspection,
it becomes evident that the work either does not evaluate FIP, or
the context of the citation is, in fact, referring to
Winograd's more-popular minimal filtering algorithms for convolutional neural
networks \winoConv \textbf{rather than} Winograd's inner-product algorithm
\cite{wino}, on which our work is based. Furthermore, the work from
Bravo \ea
\cite{mm_fpga_nogain} reports no clear improvements from their FIP
implementation, and all three other prior works on FIP architectures
\cite{wino_fir} \cite{wino_sys_1level} \cite{stanford} are purely design
proposals and do not provide empirical insight into its trade-offs.

The benefits of FIP in hardware rely on the premise that the hardware footprint
of adders are cheaper than that of multipliers.
Since the arithmetic complexity of
fixed-point multipliers scales quadratically with the input bitwidth compared to
linearly for adders, this premise has been shown to hold true for fixed-point
data types \multComplexity.
However, the exact benefits in performance/resources/clock
frequency when applying FIP to ML acceleration are still not immediately clear,
and
finding this out requires a comparison of architecture implementations that use
multipliers/accumulators of the specific data sizes/types commonly used in
machine learning accelerators.

\subsection{Traditional Inner Product (Baseline)}
Consider the matrix multiplication $\bm{C} = \bm{A} \bm{B}$ for $\bm{A}$ of size
$M \times K$ and $\bm{B}$ of size $K \times N$. Using the traditional inner product,
$\bm{C}$ is calculated from $MNK$ multiplications and $MN\lb K-1 \rb$ additions,
where each element $c_{i,j}$ of $\bm{C}$ is calculated as follows:
\begin{align}
    \label{baseline}
    c_{i,j} = \sum_{k=1}^{K} a_{i,k} b_{k,j} \,.
\end{align}

When accelerated on systolic array architectures, the majority of the
computational workload in ML models is mapped to matrix multiplication, i.e.
\eq{\ref{baseline}}, and as can be seen, the operations in this equation are a
series of multiply-accumulate operations. For all ML accelerators,
unless additional algebraic innovations are
used, the throughput is ultimately limited by the maximum number of
multiply-accumulate
operations from \eq{\ref{baseline}} that can be performed per clock cycle.
Therefore, an accelerator's computational roof is limited by the number of
multiplier and adder units it contains.

However, the arithmetic complexity of
fixed-point multipliers scales quadratically with the input bitwidth compared to
linearly for adders, causing the hardware footprint of fixed-point multipliers
to
dominate that of adders in the general case \multComplexity.
Therefore, multipliers or MAC units are commonly the area-dominant computational
resources in ML accelerators \multDominant, and an accelerator's throughput can
be directly limited by how many multipliers its hardware budget can afford. Due
to this, surpassing this theoretical performance per multiplier limit has been
focused on recently with Winograd's minimal filtering algorithms applied to
convolutional neural networks (CNN)s \winoConv \cite{liu2021winocnn}
\cite{yepez2020stride}.

In this work, we detail an \textbf{under-explored} and extremely competitive
alternative to these prior works for surpassing this performance wall and
evaluate in-depth how it compares to previous state-of-the-art solutions in
\secn \ref{sec:tables}.

%% file: algorithms.tex
\section{Fast Inner-Product Algorithms}
\label{sec:algs}
\subsection{Fast Inner Product (FIP)}
\label{sec:fip}
In FIP \cite{wino}, each element $c_{i,j}$ of $\bm{C}$ is calculated as follows:
\begin{align}\label{fip}
    c_{i,j} = \sum_{k=1}^{K/2} \lb  a_{i,2k-1} + b_{2k,j} \rb  \lb  a_{i,2k} +
    b_{2k-1,j} \rb  - \alpha_i - \beta_j \,,
\end{align}
where:
\begin{align}
    \label{eq:alpha}
    \alpha_i = \sum_{j=1}^{K/2}a_{i,2j-1} \cdot a_{i,2j}
\end{align}
\begin{align}
\label{eq:beta}
    \beta_j = \sum_{i=1}^{K/2}b_{2i-1,j} \cdot b_{2i,j} \,.
\end{align}
Calculating $\bm{C}$ then requires:
\begin{align}
\frac {MNK + MK + NK} {2}
\end{align}
multiplications for even $K$ and:
\begin{align}
  \label{fip:adds}
  \frac{3 \cdot MNK + MK + NK} {2} - MN - M - N
\end{align}
additions.
This means that an accelerator performing this equation instead of
\eq{\ref{baseline}} can trade nearly half of its multipliers for low-bitwidth
adders
while achieving the same throughput. Since the hardware footprint of fixed-point
multipliers
dominates that of adders \multComplexity, FIP can theoretically
significantly improve the throughput per compute area \roof of an accelerator.

%%%%%%%%%%%%%%%%%%%%%%%%%%%%%%%%%%%%%%%%%%%%%%%%%%%%%%%%%%%%%%%%%%%%%%%%%%%%%%
\subsection{Free-pipeline Fast Inner Product (FFIP)}
\label{sec:wino-improved}
This section introduces a new inner-product algorithm referred to as the
Free-pipeline Fast Inner Product (FFIP) that, in hardware, achieves
the same near 2\x reduction in required MAC units as FIP \cite{wino} while also
addressing a key weakness of FIP by inherently improving the clock frequency
and, as a consequence, overall throughput of an accelerator for a similar
hardware cost compared to FIP.

In FFIP, each element $c_{i, j}$ of $\bm{C}$ is calculated as follows:
\begin{align}
\label{ffip}
  c_{i,j} = \sum_{k=1}^{K/2}g^{(j)}_{i,2k-1} \cdot g^{(j)}_{i,2k} - \alpha_i - \beta_j \,,
\end{align}
where $\alpha$ and $\beta$ are the same as in \equations{\ref{eq:alpha}} and
\eqPost{\ref{eq:beta}}, and:
\begin{subequations}
\label{ffip:add1}
\begin{align}
  \label{ffip:add1-a}
  g^{(j)}_{i,2k-1} &=
  a_{i,2k} + y_{2k-1,j}   &\text{for } j = 1\\
  \label{ffip:add1-b}
  g^{(j)}_{i,2k} &=
  a_{i,2k-1} + y_{2k,j}   &\text{for } j = 1\\
  \label{ffip:add1-c}
  g^{(j)}_{i,k}      &= g^{(j-1)}_{i,k} + y_{k,j} &\text{ for } j > 1
\end{align}
\end{subequations}
\begin{align}
  \label{ffip:add3}
y_{i,j} &&=&&
  \begin{cases}
 b_{ij} & \text{for } j = 1\\
 b_{ij} - b_{i,j-1} & \text{for } j > 1 \,.
 \end{cases}
\end{align}

The extra subtractions in \eq{\ref{ffip:add3}} require $\Theta\lb NK\rb$ operations
to calculate, which is negligible since the complexity is dominated by the
overall $\Theta\lb MNK \rb$ from \eq{\ref{ffip}}. Similarly to \eq{\ref{fip}},
\eq{\ref{ffip}} also results in $\left(MNK + MK + NK\right)/2$ multiplications
for even $K$ and the same number of additions as in \eq{\ref{fip:adds}} as well.

The resulting terms being multiplied in \eq{\ref{ffip}} are identical to those
being multiplied in \eq{\ref{fip}}. The difference between \equations{\ref{fip}}
and \eqPost{\ref{ffip}} lies in the terms being fed to the addition in
\eq{\ref{ffip:add1-c}}. This allows the addition output $g^{(j)}_{i,k}$ in
\eq{\ref{ffip:add1-c}} to be passed directly into the adjacent adder for
$g^{(j+1)}_{i,k}$, which we later show improves the inner-product's hardware
frequency at essentially no cost.
Additionally, although each $g^{(j-1)}_{i,k}$ takes $j-1$ iterations to
form before
being passed to the next $g^{(j)}_{i,k}$, \secn \ref{sec:arch} shows how
these iterations for all $g$ terms in the systolic array are each being performed
in parallel in a pipeline every clock cycle. Therefore, after an
initial latency that is negligible relative to the entire execution
time, every matrix product will thereafter be produced in a pipeline at a
consistent throughput.

\subsubsection{Proof}
For $j = 1$, one can substitute the $g^{(j)}_{i,k}$ and $y_{i,j}$ terms for
their values specified in \equations{\ref{ffip:add1-a}},
\eqPost{\ref{ffip:add1-b}}, and \eqPost{\ref{ffip:add3}}, and observe the same
terms as in \eq{\ref{fip}}. For $j > 1$, \eq{\ref{ffip}} is proven by first
showing that \eq{\ref{fip}} can be rewritten as:
\begin{align}
\begin{split}
\label{proof-0}
    c_{i,j} = \sum_{k=1}^{K/2} &
    \left(
          \left[a_{i,2k} + b_{2k-1,j-1} \right]
        + \left[b_{2k-1,j} - b_{2k-1,j-1} \right]
    \right)\\
    &\left(
          \left[a_{i,2k-1} + b_{2k,j-1} \right]
        + \left[b_{2k,j} - b_{2k,j-1}   \right]
    \right)
     - \alpha_i - \beta_j\\
= \sum_{k=1}^{K/2} &
  \lb h^{(j-1)}_{i,2k-1} + y_{2k-1,j}\rb 
  \lb h^{(j-1)}_{i,2k} + y_{2k,j}\rb 
  - \alpha_i - \beta_j \,,
\end{split}
\end{align}
where:
\begin{subequations}
\begin{align}
h^{(j-1)}_{i,2k-1} &= a_{i,2k} + b_{2k-1,j-1}\\
h^{(j-1)}_{i,2k}   &= a_{i,2k-1} + b_{2k,j-1} \,,
\end{align}
\end{subequations}
and therefore:
\begin{subequations}
\begin{align}
   h^{(j)}_{i,2k-1} &= a_{i,2k} + b_{2k-1,j}\\
   h^{(j)}_{i,2k} &= a_{i,2k-1} + b_{2k,j} \,.
\end{align}
\end{subequations}
Now, $h^{(j)}_{i,2k-1}$ and $h^{(j)}_{i,2k}$ can also be rewritten as:
\begin{subequations}
\begin{align}
\begin{split}
   h^{(j)}_{i,2k-1}
   &= \lb a_{i,2k} + b_{2k-1,j-1} \rb  +\lb b_{2k-1,j} - b_{2k-1,j-1}\rb \\
   &= h^{(j-1)}_{i,2k-1} + y_{2k-1,j}\\
\end{split}
\end{align}
\begin{align}
\begin{split}
   h^{(j)}_{i,2k}
   &= \lb a_{i,2k-1} + b_{2k,j-1} \rb  +\lb b_{2k,j} - b_{2k,j-1}\rb \\
   &= h^{(j-1)}_{i,2k} + y_{2k,j} \,.
\end{split}
\end{align}
\end{subequations}
Therefore, $h^{(j)}_{i,k} = h^{(j-1)}_{i,k} + y_{k,j}$ and holds the equivalent
property as \eq{\ref{ffip:add1-c}}, showing that $h^{(j)}_{i,k} = g^{(j)}_{i,k}$
and that \eq{\ref{proof-0}}, which is equivalent to \eq{\ref{fip}}, is
equivalent to:
\begin{align}
\begin{split}
    c_{i,j}
= \sum_{k=1}^{K/2} &
  \lb g^{(j-1)}_{i,2k-1} + y_{2k-1,j}\rb 
  \lb g^{(j-1)}_{i,2k} + y_{2k,j}\rb 
  - \alpha_i - \beta_j\\
= \sum_{k=1}^{K/2} &
  g^{(j)}_{i,2k-1} \cdot g^{(j)}_{i,2k}
  - \alpha_i - \beta_j \,,
\end{split}
\end{align}
which is equivalent to \eq{\ref{ffip}}.

\subsection{Machine-Learning-Specific Optimizations}
\label{sec:alg-ml-opt}
For the application of FIP and FFIP to machine learning, $a$ values represent
the layer inputs of a DNN layer, $b$ values represent the layer weights, and $c$
values correspond to the layer's output values. The $\beta$ term in the
FIP and FFIP algorithms is a function of the weights. Hence, it remains constant
over each inference and can be pre-computed after training. The $\beta$
terms can also be incorporated into the biases, allowing the $\beta$ subtraction in
\eq{\ref{fip}} to be performed during the bias addition stage at no extra
cost. Each $\beta_j$ term is then subtracted from $bias_j$ as follows:
\begin{align}
    bias_j = bias_j - \beta_j \,,
\end{align}
where each $j$ represents a layer output channel. This allows the computation
from \eq{\ref{ffip}} to be reduced to:
\begin{align}
  c^'_{i,j} = \sum_{k=1}^{K/2}g^{(j)}_{i,2k-1} \cdot g^{(j)}_{i,2k} - \alpha_i \,,
\end{align}
where $c^'$ is a layer output before biasing and activation.

Finally, the $y_{i, j}$ values in FFIP from \eq{\ref{ffip:add3}} are also a
function of the layer weights, and can also be pre-computed after training.

%% file: architectures.tex
%%%%%%%%%%%%%%%%%%%%%%%%%%%%%%%%%%%%%%%%%%%%%%%%%%%%%%%%%%%%%%%%%%%%%%%%%%%%%%%
\section{Fast Inner-Product Architectures}
\label{sec:arch}
\subsection{Definitions}
\begin{itemize}
  \item $w$: The bitwidth that the weight and activations are quantized to.
  \item $d$: A bitwidth increase applied to certain locations in
    the FIP and FFIP PE datapaths, where $d=1$ if $a$ and $b$ are both signed or
    both unsigned, and $d=2$ if either $a$ or $b$ is signed while the other is
    unsigned.
  \item $X$, $Y$: The width and height of an MXU, respectively, in effective
    number of MAC units. For a \tip MXU, this refers to the the actual width and
    height in number of MAC units. For FIP and FFIP MXUs, $X$ and $Y$ refer to
    the MAC width and height required to achieve the same computational
    power if implemented on a \tip MXU. However, for FIP and FFIP, the number of
    actual
    instantiated MAC columns is $X/2$, and the number of actual instantiated MAC
    rows is $Y+1$ (where the extra row is for calculating the $\alpha$ terms as
    explained
    in \secn \ref{sec:mxu}).
\end{itemize}
\subsection{Processing Element (PE) Architectures}
\label{sec:pe}
\begin{figure}[t]
    \centering
   \includegraphics[scale=1]{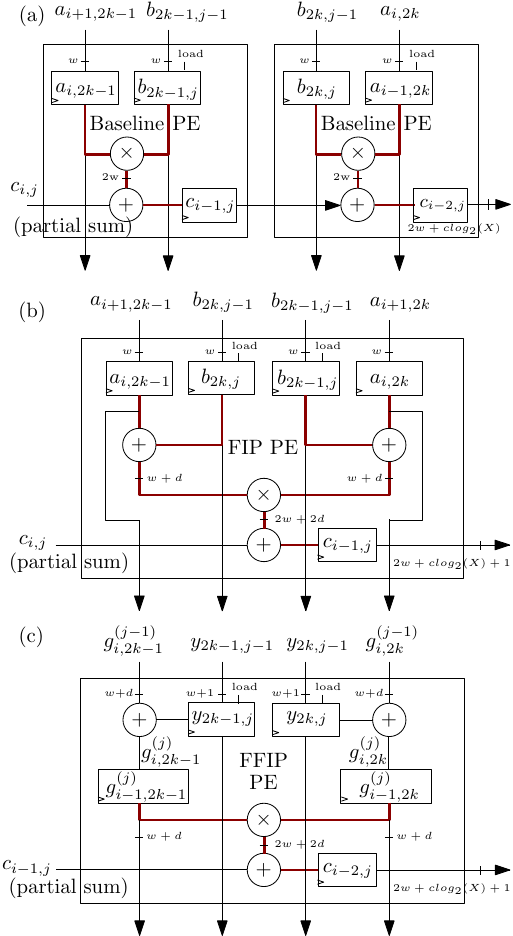}
   \caption{The PE architectures for implementing the (a) \tip, (b)
     FIP, and (c) FFIP algorithms in hardware. The FIP and FFIP PEs shown in (b)
     and (c) each individually provide the same effective computational power as
     the two baseline PEs shown in (a) combined which implement the baseline
     inner product as in existing systolic array ML accelerators. Critical paths
     are highlighted.
   } \label{fig:all-pe}
\end{figure}
\Figure \ref{fig:all-pe}a shows two traditional PEs which implement the \tip
inner-product algorithm as in existing state-of-the-art systolic
array ML accelerator architectures \citeSysArr. \Figures \ref{fig:all-pe}b and
\ref{fig:all-pe}c show PE architectures for implementing the FIP and FFIP
algorithms in an MXU hardware architecture, respectively.
The FIP and FFIP PEs shown in
\Figures \ref{fig:all-pe}b and \ref{fig:all-pe}c
each individually provide the same effective
computational power as the combined computational power of the two baseline PEs
shown in \Figure \ref{fig:all-pe}a which implement the baseline inner product.

As can be seen, compared to the baseline PEs,
the FIP and FFIP PEs therefore provide the same computational power with half
the number of multipliers, where the removed multipliers are instead traded for
cheap additions on $w$ or
$w+d$ bits. Furthermore, half of the accumulations
on $2w+clog_2(X)$ bits are also traded for lower-bitwidth additions on $w$ or
$w+d$ bits. Finally, one last benefit is that the MXUs using FIP and FFIP PEs
have a latency that is $X/2$ fewer clock cycles than a \tip MXU.

The resulting terms being multiplied in \Figure \ref{fig:all-pe}c are the
same as in \Figure \ref{fig:all-pe}b, however, the addition outputs
$g^{(j)}_{i,k}$ (the elements $g^{(j)}_{i-1,2k-1}$ and $g^{(j)}_{i-1,2k}$ in \Figure
\ref{fig:all-pe}c) can now be passed directly into the adder inputs of the
next adjacent PE below. This is beneficial because the addition outputs
$g^{(j)}_{i,k}$ stored in registers now serve the dual purpose of:
\begin{enumerate}
  \item Acting as pipelining registers for the addition outputs in
    \eq{\ref{ffip:add1}}, allowing the $g^{(j)}_{i,k}$ terms to be buffered
    directly before the multiplication, and
\item Serving as systolic buffers to store the $g^{(j)}_{i,k}$ input for the
  adjacent PE below.
\end{enumerate}
In contrast, the FIP algorithm from \eq{\ref{fip}} requires the inputs ($a$ and
$b$ in this case) to be stored in systolic buffers prior to addition before
passing them to the adjacent PEs below, but doing so does not serve the dual
purpose of also acting as pipelining registers before the multiplication.
Therefore, the FFIP algorithm inherently results in a higher maximum frequency
for a similar hardware cost.

However, in order to facilitate the usage of the FFIP PEs and gain their
benefits, it is
mandatory to fundamentally change the systolic array data flow of the original
FIP algorithm in a non-obvious way requiring a high-level mathematical
understanding of the algebra being performed, and is something that goes beyond
the capabilities of today's CAD tools. Our FFIP algorithm defined in \secn
\ref{sec:wino-improved} provides these required high-level mathematical changes.

\subsubsection{FFIP PE versus FIP PE with Additional Registers}
\label{sec:regs}
After investigating and implementing the FIP algorithm in hardware, a key
downside we found was that the FIP implementation results in a reduction in
clock frequency and, as a consequence, throughput in the resulting architecture.
As illustrated in \Figure \ref{fig:all-pe}b,
this reduction in clock frequency comes from the fact that the longest path
between registers that the logic in the FIP PE has to traverse through is across
two adders and one multiplier rather than just one adder and multiplier.
While additional registers could be placed at the FIP PE multiplier inputs to
match the critical path of the FFIP PE, this adds a significant
cost in registers in the PE array compared to using FFIP. The
register requirements for each FIP PE would then increase from
\eq{\ref{fip0-regs}} to \eq{\ref{fip1-regs}}:
\begin{align}
  \label{fip0-regs}
  4w + \lb 2w + clog_2(X) + 1 \rb = 6w + clog_2(X) + 1
\end{align}
\begin{subequations}
\label{fip1-regs}
\begin{align}
2(w+d) + \lb 6w + clog_2(X) + 1 \rb \\ =
8w + 2d + clog_2(X) + 1 \, ,
\end{align}
\end{subequations}
where \eq{\ref{fip1-regs}} is derived by adding to \eq{\ref{fip0-regs}} the
register sizes that would be required to buffer the multiplier inputs.
In contrast, the register requirements for each FFIP PE is:
\begin{subequations}
\label{ffip-regs}
\begin{align}
2\lb w+d\rb + 2\lb w+1\rb + \lb 2w + clog_2(X) + 1\rb \\=
6w + 2d + clog_2(X) + 3 \,.
\end{align}
\end{subequations}
\begin{figure}[t]
  \centering
  \hspace*{-0.35cm}
  \includegraphics[scale=0.52]{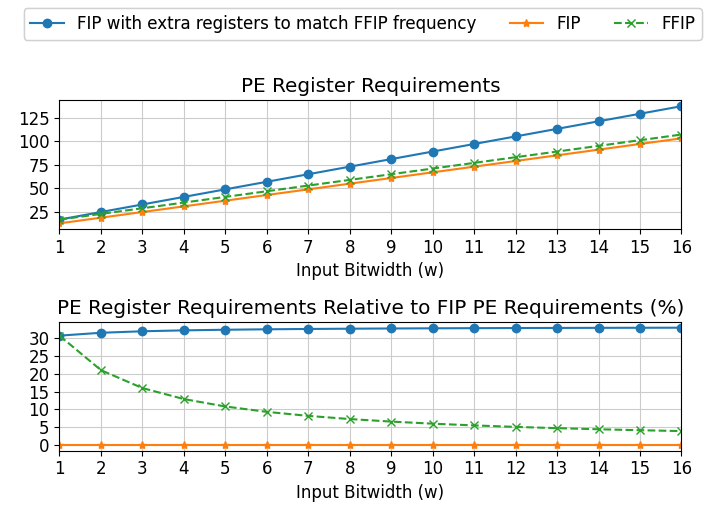}
  \vspace*{-0.7cm}
  \caption{PE register requirements at different $w$ bitwidths for the FIP PE,
    FFIP PE, and FIP PE with extra registers before the multiplier to match the
    frequency of the FFIP PEs. Values are calculated using
    \equations{\ref{fip0-regs}} - \eqPost{\ref{ffip-regs}} for $X = 64$ and $d =
    1$.} \label{fig:regs}
\end{figure}
\Figure \ref{fig:regs} shows that the FFIP register overhead defined in
\eq{\ref{ffip-regs}} starts to increase more rapidly for bitwidths
below 4, which may make it less desirable to use below that width.
However, outside of this range, FFIP register requirements are significantly
less than an FIP PE with extra registers added to match the critical
path/frequency of an FFIP PE.

Finally, while there will also be other registers to consider in an accelerator
outside of the PEs in the datapath and control logic, these register
utilizations are highly implementation-specific and will vary depending on the
details of the system implementation used to house the systolic array.
Furthermore, as the systolic array size increases, the number of PEs increase
quadratically and their total register contributions determined by
\equations{\ref{fip0-regs}}, \eqPost{\ref{fip1-regs}}, or
\eqPost{\ref{ffip-regs}} will dominate other registers outside of the systolic
array.

\subsection{Matrix Multiplication Unit (MXU) Architectures}
\label{sec:mxu}
\begin{figure}[]
    \centering
    \includegraphics[scale=1.17]{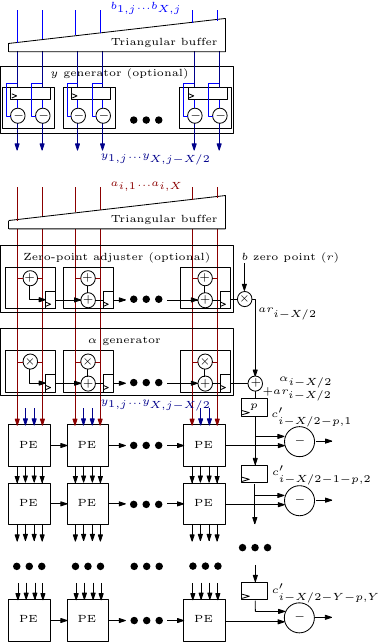}
    \caption{The FFIP MXU architecture.
      The FIP MXU is the same except that FIP PEs are used instead and the $y$
      generator block is not present, and $b$ inputs are passed in instead of
      $y$
      inputs. The $\alpha$ terms are calculated and subtracted as shown by first
      passing the $a$ inputs through an additional row of MAC units before
      they enter the rest of the MXU.}
    \label{fig:mxu}
\end{figure}

\Figure \ref{fig:mxu} shows how the PEs are laid out into an MXU architecture.
The suggested MXU input buffers shown are triangular-shaped
register arrays containing $X$ shift registers of varying depths, with each
shift
register $SR_k$ loading one $a_{i,k}$ or $b_{k,j}/y_{k,j}$ value per clock
cycle. The depth of
each $SR_k$ is $\ceil*{k / 2}$ for the FIP and FFIP MXUs, and $k$ for
the \tip MXU. In order to perform general matrix multiplication (GEMM) on a
MXU, the input matrices are divided into tiles fed to the MXU one-by-one.
Following each tile multiplication, the partial tile products are accumulated
outside of the MXU to generate each final matrix product tile. Prior to each
tile multiplication, a $b$/$y$ tile is loaded
into the MXU. It then remains in place as the $a$/$g$ tile flows through the MXU
producing the tile product, during which a new $a_{i}$ vector is fed into the
MXU each clock cycle. Additionally, to hide the latency of loading $b$/$y$
tiles,
we suggest the MXU contains an extra $b$/$y$ tile buffer which
loads the next tile as the current tile is being multiplied.

\subsection{Machine-Learning-Specific Optimizations}
\label{sec:ml-mxu}
In quantized ML inference, the weights and activations can each be quantized to
signed or unsigned integers by choosing different zero points to quantize the
values with \cite{int_quant}. However, for the FIP and FFIP architectures it is
beneficial to quantize the weights and activations such that they are
\textit{both} represented on either signed or unsigned integers. Otherwise,
representing one on signed and the other on unsigned adds a penalty in the
hardware footprint of the FIP and FFIP PEs because then, due to the possible
range of the result, $d$ must equal 2 instead of $1$,
and the sum of $a$ and $b$ elements must then be represented on $w+2$ bits
rather than the $w+1$ bits needed if they were both signed or both unsigned.
This would lead to extra register requirements in the FFIP PE and cause
multiplication on $w+2$-bit inputs rather than $w+1$ bits for both the FIP and
FFIP PEs.

Furthermore, if the employed quantization scheme uses nonzero zero points for
the weights, the contributions of the zero points in the GEMM product must then
be separately calculated and subtracted from the result to eliminate their
contribution (for any architecture, not just FIP and FFIP) \cite{int_quant}. We
provide a solution to mitigate this penalty in the FIP and FFIP architectures
when layer-wise zero points are used on the weights.
We show how to pass some of these extra required computations
into the pre-existing $\alpha$ generator logic as shown in \Figure \ref{fig:mxu}. To
help illustrate how this works, consider that the weight zero points can be
represented by a constant matrix $\bm{R}$ being added to the weights where each
element has a constant value. This results in the MXU performing the following
computation:
\begin{align}
  \bm{A} \left( \bm{B} + \bm{R} \right) = \bm{A}\bm{B} + \bm{A}\bm{R} \,.
\end{align}
Therefore, in order to eliminate the contribution of the constant matrix
$\bm{R}$ in the matrix product, the $\bm{A}\bm{R}$ product must be subtracted
from the MXU output. In order to do this and combine its subtraction with the
existing $\alpha$ generator logic, we provide the \textit{zero-point adjuster} block
shown in \Figure \ref{fig:mxu}, which calculates the $\bm{A}\bm{R}$ elements
using only one multiplier, and show how to efficiently combine its
$\bm{A}\bm{R}$ output elements with the $\alpha$ elements. This way, both the
$\bm{A}\bm{R}$ and $\alpha$ values are subtracted from the MXU output vectors at the same
time and an additional hardware subtraction vector dedicated to subtracting only
the $\bm{A}\bm{R}$ elements is not required on the MXU output.

Finally, as discussed in \secn \ref{sec:alg-ml-opt}, the $\beta$ terms do not need
to be calculated as they do for $\alpha$ since they can be pre-computed and added to
the biases. Furthermore, the $y$ values can either be calculated in real time
with the $y$ generator as shown in \Figure \ref{fig:mxu}, or they can be
pre-computed at the cost of storing them in 1 extra bit in memory.

%% file: system.tex
\section{System Architecture}
\label{sec:system}
\begin{figure}[]
    \centering
    \includegraphics[scale=.7]{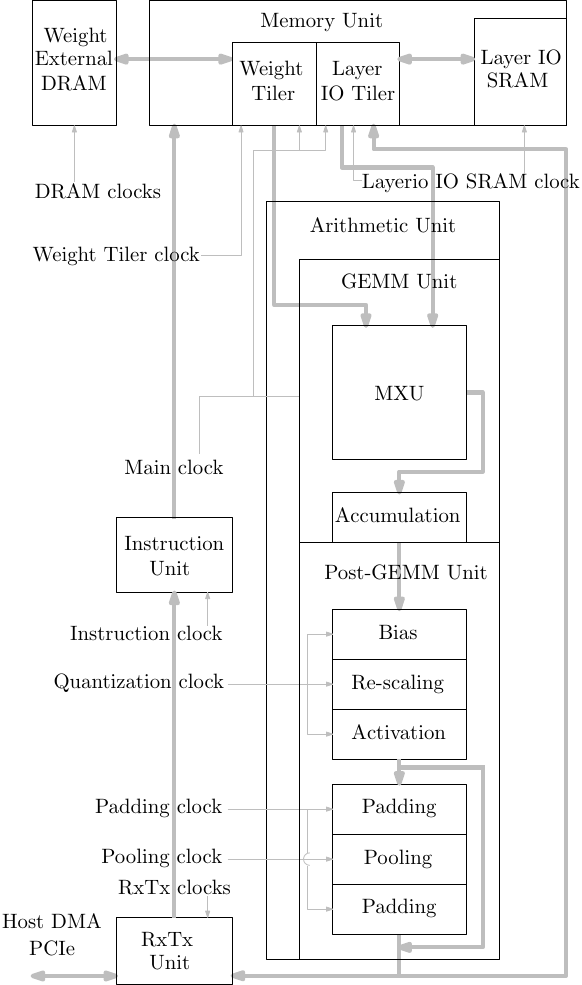}
    \caption{The example accelerator system design used to host and evaluate the
      baseline, FIP, and FFIP MXUs when overlaid on top of systems based on the
      most efficient systolic array accelerators used in
      practice, e.g., the TPU \cite{tpu-v1}\cite{tpu-v3}\cite{tpu-v4}.}
    \label{fig:sys}
\end{figure}
The key contributions of this work are the proposed FFIP algorithm and
architectures and their application to ML acceleration, which improve the
under-explored FIP \cite{wino} that has never before been applied to ML
acceleration. However, in order to ensure that there are no unseen side-effects,
it is important to integrate the FFIP architecture in a full-system
implementation to fully assess the end benefits at the application level.
Furthermore, our goal is to validate the benefits of FFIP when overlaid on top
of systems based on the most efficient systolic array accelerators used in
practice. Therefore, the high-level organization of our system design shown in
\Figure \ref{fig:sys} \textit{intentionally} shares similarities with the TPUv1
\cite{tpu-v1} and some similarities with the TPUv2-TPUv4 as well \cite{tpu-v3}
\cite{tpu-v4}. Based on this, we provide here the system-level implementation we
used for the sake of completeness and for better understanding of the context in
which our main contributions have been validated.

\subsection{Memory Subsystem}
\begin{figure}[]
  \hspace*{-0.7cm}
  \centering
  \includegraphics[scale=1]{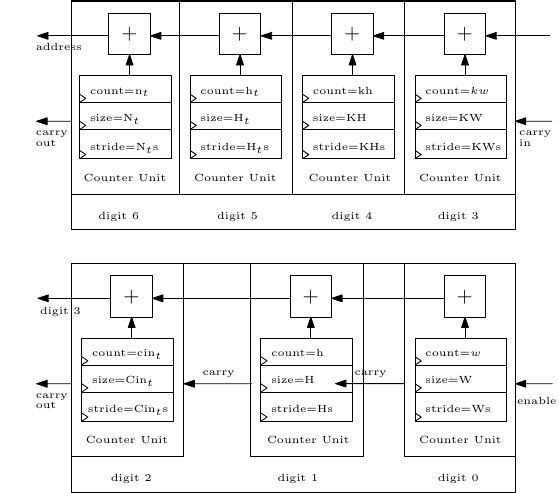}
  \caption{Layer IO memory access counters for performing in-place mapping
    of 2-D convolution to GEMM.}
  \label{fig:mem-counters}
\end{figure}

The memory accesses are controlled by multi-digit counters (also referred to as
tilers) shown in \Figure
\ref{fig:mem-counters} containing programmable digit sizes and strides which
calculate GEMM read/write patterns, as well as map two-dimensional convolution
to matrix multiplication to perform Algorithm \ref{alg:mem} for any matrix or
convolution window sizes
without requiring a stand-alone memory remapping stage.
The size and strides for
each layer are calculated offline once per neural network, and then can be
re-used for all inferences of the same neural network after that where they are
updated in the memory tilers in real time between each layer.
The tilers allow the \Mem and external dynamic random-access memory (DRAM) to be
interfaced from the
Arithmetic Unit using simple first-in first-out (FIFO) interfaces.

\begin{figure}[]
  \centering
  \includegraphics[scale=.5]{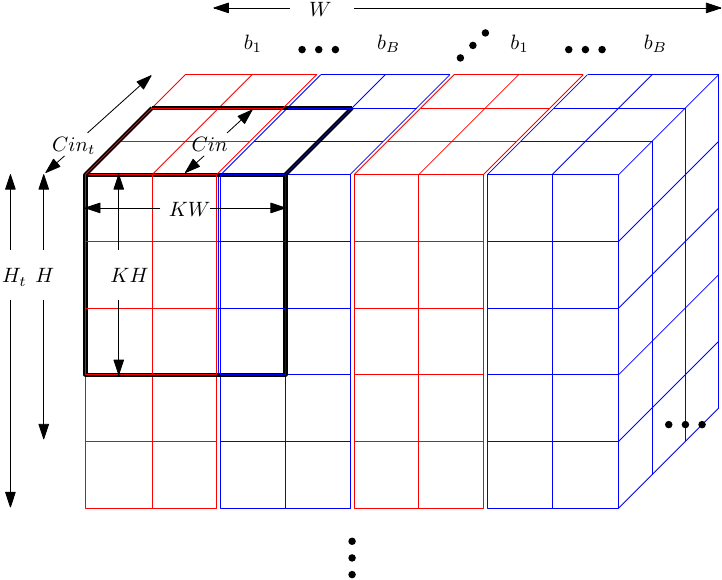}
  \caption{Layer IO memory access and blocking scheme to partition the memory
    and perform in-place mapping of two-dimensional convolution to GEMM,
    provided for context.}
  \label{fig:mem-blocks}
\end{figure}

\begin{algorithm}[t!]
  \caption{Layer IO memory access pattern implemented in the counters from
    \Figure \ref{fig:mem-counters} for in-place mapping of 2-D convolution to
    GEMM, provided for the sake of completeness.}
  \label{alg:mem}
  \begin{algorithmic}[1]
    \NoDo
        {\small
    \For {$n_t$ = 0; $n_t <N_t \cdot N_ts$; $n_t$ += $N_ts$}
    \For {$h_t$ = 0; $h_t <H_t \cdot H_ts$; $h_t$ += $H_ts$}
    \For {$kh$ = 0; $kh <KH \cdot KHs$; $kh$ += $KHs$}
    \For {$kw$ = 0; $kw <KW \cdot KWs$; $kw$ += $KWs$}
    \For {$cin_t$ = 0; $cin_t$<$Cin_t$$\cdot$$Cin_ts$; $Cin_t$ += $Cin_ts$}
    \For {$h$ = 0; $h <H \cdot Hs$; $h$ += $Hs$}
    \For {$w$ = 0; $w<W\cdot Ws$; $w$ += $Ws$}
    \State $k_{\text{offset}} = kh + kw + cin_t$
    \State $m_{\text{offset}} = h_t + h + w$
    \State address $ = m_{\text{offset}} + k_{\text{offset}}$
    \EndFor
    \EndFor
    \EndFor
    \EndFor
    \EndFor
    \EndFor
    \EndFor
}
  \end{algorithmic}
\end{algorithm}

The convolutional layer inputs consist of three dimensions, depth $Cin$, height
$H$, and width $W$. The $Cin$ and $H$ dimensions are each split into two
separate dimensions to divide them into tiles, and a dimension with the
$_t$ subscript
denotes the number of tiles in that dimension (e.g. the $H$ dimension is split
into $H_t$ tiles each of a new smaller size $H$). Consider that $KH$ and $KW$
stand for the height and width of the kernel size, that a dimension suffixed
with $s$ represents the
dimension's stride (e.g. $Ws$ is the stride for the $W$ dimension), and $N_t$
stands for the number of tile columns in the $B$/weight matrix. The convolution
is then mapped in-place to GEMM as shown in Algorithm \ref{alg:mem} where all of
the layer input dimensions are mapped to tiled matrices by mapping each
dimension to either a $K$ or $M$ dimension. Similarly, the weight dimensions are
tiled and mapped to either the $K$ or $N$ dimensions for GEMM. Each memory
location contains $X$ elements along the $Cin$ dimension, where $X$ is the MXU
width.

\subsubsection{Memory Subsystem Timing Optimizations}
It was initially found that the maximum frequency of the memory access
control counters was lower than the GEMM Unit clock frequency, which would lower
the throughput of the entire accelerator and restrict the full frequency
advantages of FFIP from being properly evaluated. To resolve this, we
partitioned the Layer IO memory into submemories each containing a block of the
total memory and separate memory tiler controllers. This allowed each memory
block to be accessed concurrently at a slower clock speed and for their read
data to then be accessed in an interleaved manner by the main clock.

To do this, we derived a memory partitioning scheme for dividing the
convolutional layer inputs into blocks in a way that still supports in-place
remapping of two-dimensional convolution to GEMM as described above. The
partitioning scheme supports dividing the memory into $B$ blocks for any $B$
that is a power of 2, allowing the memory tiler controllers to run at a
frequency of 1/$B$ times the main clock frequency. In our results, we use $B=2$.
The $W$ dimension is divided into different blocks as show in \Figure
\ref{fig:mem-blocks}, where each $W$ slice is $Ws$ elements wide.

One remaining issue, however, is that when the $KW$ dimension increments to a
certain range, the submemory may need to access elements from a block that it
does not contain. For example, consider the case in \Figure \ref{fig:mem-blocks}
where $kh = kw = 3$, $Hs = Ws = 2$, and $B = 2$. When $kw \in \{1, 2\}$ then
element 1 in block 1 will be accessed by the main clock for $w = 1$
followed by element 1 in block 2 for $w = 2$. However, when $kw = 3$ then
the block 1 submemory should require access starting with an element from a
block that it does not contain. To resolve this, when $kw = 3$ then block 2 will
be accessed first by the main memory for $w = 1$, followed by the next
block 1 slice for $w = 2$. In other words, the next memory elements are instead
taken from the adjacent submemory, the $kw$ and $w$ incrementer digits are
adjusted in each submemory, and the interleaving order in which the submemories
are accessed from the main clock is modified.

We also run the weight memory control logic at a fraction of the main clock
speed by accessing the memory in bursts and thus requiring memory control to be
communicated to the DRAM controller at an infrequent rate relative to the main
clock speed. The external DRAM memory is used only for storing the weights, and
the layer inputs/outputs always stay in on-chip memory. This allows the device's
external memory bandwidth to rarely be a bottleneck for the implementations
evaluated in our experimental setup discussed in \secn \ref{results}.

\subsection{MXU Timing Optimizations}
\begin{figure}[]
  \centering
  \includegraphics[scale=0.6]{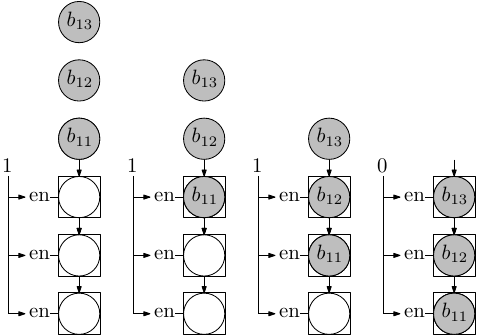}
  \caption{MXU weight column shift register logic requiring control signal to be
    connected to each element in the column.}
  \label{fig:shift-reg}
\end{figure}

\begin{figure}[]
  \centering
  \includegraphics[scale=0.6]{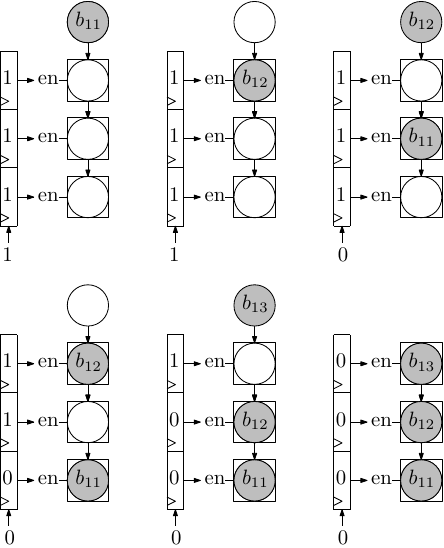}
  \caption{MXU weight column shift register logic with control connections fully
    localized to adjacent elements.}
  \label{fig:shift-reg2}
\end{figure}

As discussed in \secn \ref{sec:mxu}, prior to each tile multiplication, a
$b$/$y$ tile is loaded into the MXU. It then remains in place as the $a$/$g$
tile flows through the MXU to produce the tile product. As show in \Figure
\ref{fig:shift-reg}, our initial implementation for the mechanism which shifts
$b$/$y$ tiles into the MXU was to implement a basic one-dimensional shift
register vector to shift each column of weights into the MXU, and then have an
enable control signal that connects to each element in the vector to stop the
shifting all at once when the weight column is fully loaded. However, this
required enable control signals to be connected non-locally to every element's
shift register enable inputs without being able to locally buffer the enable
control signals close to their destinations.

In order to mitigate this, we instead designed an improved shift-register-based
mechanism shown in \Figure \ref{fig:shift-reg2} in which the control signal
connections are fully localized by being buffered directly before their
destination in each element's shift register enable input to improve the
frequency potential of the MXU. In this design, the control signal has its own
shift register mechanism separate from the datapath shift register which is
pre-loaded with 1's. To allow this mechanism, the weights are shifted in
every-other clock cycle instead of every clock cycle. Shifting in the weight
vectors more slowly like this does not affect the throughput so long as the
layer input $M_t$ tile size can usually be at least twice
as large as the $N_t$
tile size used for the weights, which was found to be true for
the models we used for evaluation.

The key contributions of this work are the proposed FFIP
algorithm and architecture and their evaluation in an ML accelerator system,
which improve the
under-explored FIP \cite{wino} that has never before been applied to ML
acceleration.
Furthermore, our goal is to evaluate FFIP when
overlaid on top of a system based on the most efficient systolic array
accelerators used in practice. Therefore,
this section expanded upon our system design shown in \Figure \ref{fig:sys}
that \textit{intentionally} shares similarities
with the TPUv1 \cite{tpu-v1} and some similarities with the
TPUv2-TPUv4 as well \cite{tpu-v3} \cite{tpu-v4},
and provides the context in
which our main contributions have been validated to better understand
their overall impact.

%% file: results.tex
\section{Results}
\label{results}
This section demonstrates the generality of the FIP-based approaches by
evaluating our example (F)FIP implementations on neural networks.
The theoretical benefits of the FIP architecture are validated
for the first time in a machine learning accelerator. Additionally, we validate
the theoretical benefits of the proposed FFIP architecture against the baseline
and FIP architectures,
and against state-of-the-art prior works. Although the theoretical
concepts presented in this work are general and applicable to both custom
integrated circuits and field-programmable gate array (FPGA) implementations,
our FFIP algorithm and architecture were validated on FPGA, and we therefore
confirm our theoretical insights by comparing the benefits of the proposed FFIP
architecture against the best-in-class of prior state-of-the-art ML accelerator
solutions that are also evaluated on FPGA. We compare designs for non-sparse
neural
network acceleration and input sizes quantized to 8 - 16 bits, which are
bitwidths commonly used in practice as they provide a balanced trade-off between
accuracy versus hardware efficiency \cite{int_quant}.

The \textbf{source code} for our accelerator implementation is available on
GitHub \cite{ffip-source} and is hand-coded in SystemVerilog and implemented to be
highly configurable. The MXU height/width can be parameterized to be any value
that is a multiple of 4. The bitwidth of the $a$ and $b$ values are also fully
parameterizable, as well as whether they are on signed versus unsigned
representations. The design performs inter-layer rescaling for quantization
\cite{int_quant} in the Post-GEMM Unit, which requires
an additional $Y$ multipliers (for all MXUs - baseline, FIP, and FFIP). The
accelerator has direct memory access (DMA) to the system memory in the host
through a Peripheral Component Interconnect Express (PCIe) 3.0
connection. We use a PCIe DMA controller and host driver provided by Xillybus
\cite{xillybus}. From the hardware end, the PCIe controller is interfaced
through FIFOs. From the software end, the driver is interfaced through shell
scripts which we interface with through Python to provide an
end-to-end framework for running ML applications on the accelerator.

Full system-level validation of the experimental accelerator system from \secn
\ref{sec:system} has been done on an Arria 10 SoC Developement Kit
\cite{sx-dev-kit} containing the \sx device by measuring model throughput in
real-time. However, this device has significantly fewer soft logic resources
compared
to the \gx used in the majority of the works we compare against on the same
FPGA
family in this section, and our memory subsystem implementation requires the
extra memory resources available in the \gx device for the 16-bit-input
architecture. Therefore, in order to provide a comparison with the 16-bit input
version of our architecture in \secn \ref{sec:tables}, and to use the same
identical FPGA from the Arria 10 family as the majority of prior works
that also use Arria 10, we generate compilation results for our
design on the \gx device for a more fair and consistent comparison in \secn
\ref{sec:tables}.
The throughput values for our designs on the \gx are calculated using
an accurate throughput estimation analysis based on our highly deterministic and
time predictable system implementation, which predicts the actual model
throughputs measured on the \sx device available to us within an error margin
of 1\% for all MXU sizes.
The models used for evaluation are AlexNet \cite{alexnet} and ResNet
\cite{resnet}.

\subsection{FFIP Compared to \TIP and FIP}
\label{ffip-vs-fip}
\begin{figure}[]
  \centering
  \hspace*{-0.35cm}
  \includegraphics[scale=0.52]{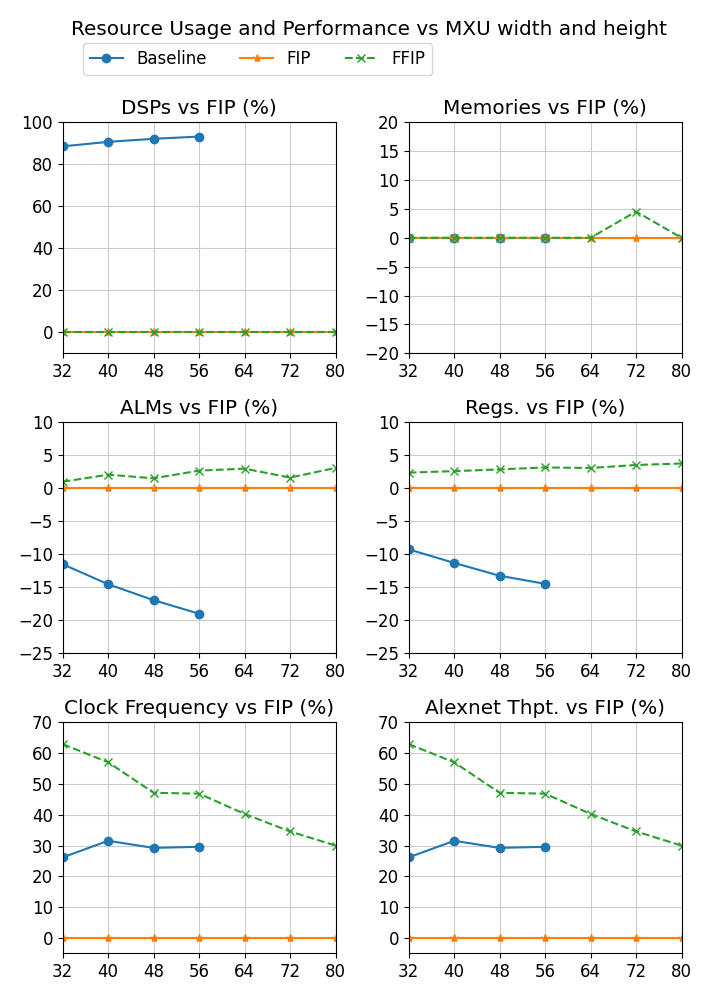}
  \caption{
      Evaluating the baseline, FIP, and FFIP MXUs at different sizes
      instantiated into an example ML accelerator system design used for
      validation, with 8-bit fixed-point inputs on an \sx FPGA.}
  \label{fig:report}
\end{figure}

The \tip, FIP, and FFIP MXUs from \secn \ref{sec:mxu} have been instantiated
inside an example accelerator system used for validation that was discussed in
\secn \ref{sec:system}. Different accelerator designs were compiled where the
system architecture remained the same, and only the MXU size and type changed.
The
MXU height/width was incremented by multiples of 8 from sizes 32 to 80, or until
the device no longer contained enough digital signal processing (DSP) units to
instantiate the design. For each
MXU size, we compiled a \tip, FIP, and FFIP MXU, except for the \tip MXU
designs above size 56\by56 which no longer fit
on the device due to the DSP resources reaching their limit.

The memory resource deviation seen in the 72\by72 FFIP MXU in \fig
\ref{fig:report} is due to the FPGA compiler placing
some of the datapath buffers outside of the MXU into memory resources.
However, due to the heuristic nature of FPGA compilers, where even
bit-equivalent designs will produce small variations in hardware utilizations
for different random seeds, this deviation is interpreted as a random
outlier and not something inherent to FFIP.

The FIP architecture uses up to
15-20\% more ALMs and registers than the baseline to implement the pre-adders
that nearly half the multipliers and accumulators are traded for as discussed in
\secn \ref{sec:pe}. However, as expected from our theoretical analysis, the FIP
architecture provides the much more
significant \textbf{near 2\x reduction in DSP units}, which are hard logic
resources used to implement MAC units in FPGAs since that operation
cannot be mapped efficiently onto the soft logic resources
\cite{fpga_nn_survey}. The clock frequency of the FIP
designs, however, is approximately 30\% lower than the clock frequency of the
\tip
designs, which consequently reduces the throughput by the same ratio.
As anticipated from our theoretical analysis, the FFIP
architecture inherently addresses this weakness of the FIP architecture
for a similar hardware cost by
maintaining the same near 2\x reduction in the number of required DSPs, however,
\textbf{FFIP improves the clock frequency by over 30\%}, and as a consequence,
the overall throughput of the accelerator is improved by the same amount
compared to FIP. Additionally, the effective size of the
largest systolic array that can be fit onto the device before maxing out the
DSPs is increased from 56\by56 PEs to 80\by80 when using FIP and FFIP, a
\textbf{2\x increase in effective number of PEs}.

Finally, while additional registers could be placed at the FIP PE multiplier
inputs to match the critical path of the FFIP PEs, this would come at a
significant extra cost in additional registers compared to using FFIP. A
detailed analysis for this is provided in \secn \ref{sec:regs}.

\subsection{FFIP Compared to the State-of-the-Art on FPGA}
Although the theoretical
concepts presented in this work are general and applicable to both custom
integrated circuits and FPGA implementations,
our FFIP algorithm and architecture were validated on FPGA, and we therefore
confirm our theoretical insights by comparing the benefits of the proposed FFIP
architecture against the best-in-class of prior ML accelerator
solutions that are also evaluated on FPGA.
\label{sec:comparison}

%%%%%%%%%%%%%%%%%%%%%%%%%%%%%%%%%%%%%%%%%%%%%%%%%%%%%%%%%%%%%%%%%%%%%%%%%%%%%%
\subsubsection{Performance Metric Definitions and Motivations}
\label{sec:metrics}
The common practice in the field for judging an ML accelerator's performance is
with the \textit{effective throughput} metric, $\thpte$, which will also be
referred to as solely the \textit{throughput}, and is defined as the number of
effective inference operations the accelerator can execute per second in real
time. It is also commonly notated as $GOPS$ or $TOPS$ to represent its value
when scaled down by one billion or one trillion, respectively. It is calculated
as follows by measuring the average number of inferences performed by the
accelerator per second, multiplied by the number of operations required to
complete each inference when using traditional algorithms for the algebra such
as \eq{\ref{baseline}}:
\begin{subequations}
\begin{align}
  throughput &= \thpte \\[5pt]
  &= \frac{\#inferences}{s} \ex \frac{\#operations}{inference} \\
  \label{alg:ops-thpt}
   &= \frac{\#operations}{s} \\
  \label{alg:add-mult-thpt}
   &\approx \frac{\#multiplications + \#additions}{s} \,.
\end{align}
\end{subequations}
While there are other operations performed other than strictly additions and
multiplications such as the \textit{greater than} operator used in Max-Pool
layers, the number of additions and multiplications performed in the majority of
ML models dominate the number of other types of operations performed
\cite{tpu-v1}, which is why \eq{\ref{alg:add-mult-thpt}} is an accurate
approximation of \eq{\ref{alg:ops-thpt}}.

Therefore, an accelerator's throughput \roof, defined as the throughput if all
instantiated compute resources are utilized in every clock cycle, is accurately
approximated by the number of instantiated multiplier and adder units multiplied
by how many executions each multiplier and adder unit can perform per second,
i.e. the accelerator's clock frequency, as follows:
\begin{align}
  \label{thptCeil0}
  \thRoof &\approx \lb \#multipliers + \#adders \rb \ex \freq \,,
\end{align}
where $\freq$ is the accelerator's clock frequency.

However, the logic complexity of fixed-point multipliers scales
quadratically with the input bitwidth compared to linearly for adders and registers, causing
the hardware footprint of multipliers to dominate that of adders and registers in
the general
case \multComplexity. Due to this, multipliers and MAC units are commonly the
area-dominant computational resources in ML accelerators \multDominant. Based on
this and the fact that \eq{\ref{baseline}} has the following property:
\begin{align}
  Baseline\spcc \#additions &\approx \#multiplications,
\end{align}
we can rewrite \eq{\ref{thptCeil0}} as follows to more directly reflect the
computational throughput \roof by basing it solely on how many of the
area-dominant computational resources, i.e. multipliers, that a device's
hardware budget can afford:
\begin{subequations}
  \label{eq:roofs-baseline-first}
\begin{align}
  Baseline\spcc &throughput\spcc roof = \thRoof\\
  &\approx \lb \#multipliers + \#adders \rb \ex \freq \\
  \label{thpt-roof-baseline}
  &\approx 2 \ex \#multipliers \ex \freq \,,
\end{align}
\end{subequations}
which can also be re-arranged to determine the roof of the following performance
per compute area metric:
\begin{subequations}
\begin{align}
  Baseline\spcc \frac{throughput} {compute\spc area} \spcc \roof &\approx
  \frac{\thpte \spcc roof} {\#multipliers}\\
  \label{area-roof-baseline}
  &\approx 2 \ex \freq \,,
\end{align}
\end{subequations}
and the roof of the following performance per compute area per clock cycle
metric:
\begin{subequations}
  \label{eq:roofs-baseline-last}
\begin{align}
  Baseline\spcc \threefrac{throughput} {compute\spc area} {clock\spc cycle}
  \spcc \roof &\approx
  \threefrac{\thpte \spcc roof} {\#multipliers} {\freq}\\
  \label{area-cc-roof-baseline}
  &\approx 2 \,.
\end{align}
\end{subequations}

However, when an accelerator instead uses FIP or FFIP for the arithmetic
(\equations{\ref{fip}} and \eqPost{\ref{ffip}}), the ratio of multiplications to
additions becomes:
\begin{align}
  (F)FIP\spcc \#additions &\approx 3\cdot\#multiplications \,.
\end{align}
This means that the device can trade nearly half of its compute-area-dominant
multipliers for low-area adders while achieving the same throughput and the same
number of inferences per second. Since the hardware footprint of the multipliers
still dominates that of the adders as explained above, the (F)FIP throughput
\roof can still be most directly reflected by basing it solely on  how many of
the area-dominant computational resources, i.e. multipliers, that a device's
hardware budget can afford:
\begin{subequations}
  \label{eq:roofs-fip-first}
\begin{align}
  (F)FIP\spcc
  &throughput\spcc roof = \thRoof\\
  &\approx \lb \#multipliers + \#adders \rb \ex \freq \\
  \label{thpt-roof-ffip}
  &\approx 4 \ex \#multipliers \ex \freq \,,
\end{align}
\end{subequations}
which can also be re-arranged to determine the roof the following performance
per compute area metric:
\begin{subequations}
\begin{align}
  (F)FIP\spcc
  \frac{throughput} {compute\spc area} \spcc \roof &\approx
  \frac{\thpte \spcc roof} {\#multipliers}\\
  \label{area-roof-ffip}
  &\approx 4 \ex \freq \,,
\end{align}
\end{subequations}
and the roof of the following performance per compute area per clock cycle
metric:
\begin{subequations}
  \label{eq:roofs-fip-last}
\begin{align}
  (F)FIP\spcc
  \threefrac{throughput} {compute\spc area} {clock\spc cycle} \spcc \roof &\approx
  \threefrac{\thpte \spcc roof} {\#multipliers} {\freq}\\
  \label{area-cc-roof-ffip}
  &\approx 4 \,.
\end{align}
\end{subequations}

Based on the above definitions and reasoning, we use the following three
performance metrics for judging prior state-of-the-art solutions against FFIP:
\begin{subequations}
\label{eq:metrics}
\begin{align}
  \label{metric:thpt}
  throughput\spcc& \spc\spc\spcc \lb GOPS \rb \spcc\spc\spcc\spc
  = \spcc\spc \thpte \ex 10^{-9}\\[5pt]
  \frac{throughput} {compute\spc area}& \lb \frac{GOPS} {multiplier} \rb
  \label{metric:thpt-area}
  \spc = \frac{\thpte \ex 10^{-9}} {\#multipliers} \\[5pt]
  \label{metric:thpt-area-cc}
  \threefrac{throughput} {compute\spc area} {clock\spc cycle}& \lb
  \threefrac{operations} {multiplier} {clock\spc cycle} \rb
  = \threefrac{\thpte} {\#multipliers} {\freq} \,.
\end{align}
\end{subequations}

For the results in \secn \ref{sec:tables},
\#multipliers is calculated as $\#DSPs$ for AMD/Xilinx FPGAs
where each DSP instantiates one 18\by27-bit multiplier \cite{amd-dsp}, and
$\#DSPs \times 2$ for Intel/Altera FPGAs where each DSP instantiates two 18\by19-bit
multipliers \cite{intel-dsp}. The only exception to this is for the works using
Intel/Altera FPGAs from Liu \ea \cite{liu2022toward} and Fan \ea
\cite{fan2022fpga}, where two 8-bit multiplications are packed onto each
18\by19-bit multiplier in
the DSPs and additional ALMs as explained in \secn
\ref{sec:tables}, and therefore $\#multipliers = \#DSP \times 4$.

We also note that, when evaluating solutions on FPGA, it is more common in the
field to measure the performance per compute area using \opsDSP rather than
\opsMult. However, we use the latter in order to make the evaluation
more generalized and to
detach it from the specific DSP architectures currently being used in modern
FPGAs. For example, the DSPs on most AMD/Xilinx FPGAs \cite{amd-dsp} contain one
fixed-point multiplier, whereas the DSPs on most Intel/Altera FPGAs
\cite{intel-dsp} contain two fixed-point multipliers, which would make
GOPS/DSP comparisons unfair when comparing designs on Intel versus AMD FPGAs,
and would decrease the generality of the comparison.

The \ops metric defined in \eq{\ref{metric:thpt}} measures the raw performance
that a solution is able to achieve on a device and is the foremost commonly used
metric for comparison in the field. We use the \opsArea and \opsAreaCC metrics
defined in \equations{\ref{metric:thpt-area}} -
\eqPost{\ref{metric:thpt-area-cc}} to help compare the scaling potential of a
design regardless of how many computational resources are available on the
device it is scaled onto, or so that prejudice is not given towards
implementations that could theoretically achieve higher throughputs but did not
scale up their design to take full advantage of all the compute resources
available on the device. The \opsAreaCC metric serves a similar purpose as
\opsArea, however, it also normalizes for clock frequency to further abstract
the performance criteria from the implementation platforms and their technology
nodes, which will have varying timing potentials, and this metric will also
remove any doubt regarding FFIP's performance improvements coming solely from
increase in clock frequency compared to the prior works. However, we still
include the \opsArea metric because a key benefit of FFIP is that it
\textit{does} inherently allow for a higher clock frequency for a similar
hardware cost compared to FIP.
%%%%%%%%%%%%%%%%%%%%%%%%%%%%%%%%%%%%%%%%%%%%%%%%%%%%%%%%%%%%%%%%%%%%%%%%%%%%%%

\subsubsection{Discussion}
\label{sec:tables}
\input{results-8b-table}
\input{results-16b-table}
\input{results-mixed}

As derived in \equations{\ref{eq:roofs-baseline-first}} -
\eqPost{\ref{eq:roofs-baseline-last}} versus
\eqPost{\ref{eq:roofs-fip-first}} -
\eqPost{\ref{eq:roofs-fip-last}}, (F)FIP increases the theoretical roof of the
performance metrics defined in \equations{\ref{metric:thpt}} -
\eqPost{\ref{metric:thpt-area-cc}} in the general case, regardless of the
implementation technology. To empirically validate this, we compare our results
in this section against the prior state-of-the-art, and since our FFIP algorithm
and generalized architecture were validated on FPGA,
we confirm our theoretical insights by comparing the benefits of the proposed
FFIP architecture against the best-in-class prior
solutions also evaluated on FPGA. We demonstrate that FFIP is extremely
competitive and
surpasses the prior works in the evaluated performance metrics.

Due to the clear advantages, increasing the theoretical limits of the
performance metrics defined in \equations{\ref{metric:thpt}} -
\eqPost{\ref{metric:thpt-area-cc}} from \secn \ref{sec:metrics} has been
focused on recently in other works as well. The works by
Yepez \ea \cite{yepez2020stride} and Jiang \ea \cite{jiang2022cpu}
do this by
exploiting Winograd's minimal filtering algorithms applied to convolutional
neural networks as shown by the work from Lavin \ea \cite{lavin_fast}. However,
as shown in Table \ref{tab:16}, the FFIP architectures surpass the prior works
in effective ability to reduce the multiplication complexity of the neural
network workloads,
shown by
comparing the \textit{operations/multiplier/clock cycle} metric defined in
\eq{\ref{metric:thpt-area-cc}}.
Furthermore,
Winograd's minimal filtering algorithms
are applicable only to convolutional layers, whereas FIP and FFIP are applicable
to all deep learning models and layer types that mainly decompose to matrix
multiplication, such as fully-connected, convolutional, recurrent, and
attention/transformer layers.
Finally, the Winograd convolution technique
\cite{lavin_fast} still results in matrix multiplication, which can therefore
still achieve further compute efficiency improvements by also
executing the
resulting matrix multiplication on a systolic array architecture housing FFIP
PEs.

Additionally, the Arria 10 DSPs \cite{intel-dsp} can each perform two
multiplications between one
18-bit and one 19-bit integer input. As demonstrated in the works by Liu \ea
\cite{liu2022toward} and Fan \ea \cite{fan2022fpga}, it is possible to pack the
multiplier inputs
such that each 18\by19-bit multiplier performs two multiplications of 6-bit
inputs.
The remaining two 2\by8-bit multiplications
and two 2\by6-bit multiplications required are then performed
and summed using extra ALMs. The works by Liu \ea \cite{liu2022toward} and Fan
\ea \cite{fan2022fpga} use this technique to improve the DSP efficiency for
8-bit inputs. However, this technique requires a noticeable extra cost in ALMs,
it is specific to the particular DSP architecture currently being used in
Intel/Altera
FPGAs \cite{intel-dsp}, and it does not work for 16-bit inputs.
We also note that the works from Liu \ea \cite{liu2022toward} and Fan \ea
\cite{fan2022fpga}
report the total resources available on the FPGA in their comparison with prior
work, however, we list their total instantiated resources that are also reported
in their work
in our comparison since it is the more common practice and
is more favourable for them to use in the comparison here.

We also note that, while our system implementation required more FPGA memory
resources than some of the other works, the FFIP architecture proposed in this
work affects only to the systolic array of an accelerator and does not consume
any of the memory resources itself. The memory subsystem design used in our
example accelerator implementation that is consuming the majority of the FPGA
memory resources was used to ensure that off-chip memory bandwidth was not a
bottleneck on the Arria 10 SoC Developement Kit available to us in order to
ensure we could properly evaluate the benefits of FFIP on that platform. FFIP
can be used in any accelerator system that uses traditional fixed-point systolic
arrays for the arithmetic without fundamentally altering the accelerator's
functionality or internal interfaces in any way, and its usage is orthogonal to
the memory subsystem being used in the accelerator.

To make comparisons as fair as possible, we aligned as many variables as
possible to the best of our ability in the comparisons in Tables
\ref{tab:first} - \ref{tab:last} in regards to making comparisons against
implementations on the same FPGA, the same input/datapath bitwidths, and the
same ML models. However, since it is not feasible in implementation time to
evaluate any one solution for all ML models on all platforms, Tables
\ref{tab:8} and \ref{tab:16} compare
prior works on the same FPGA family and the same datapath/input
bitwidths but for varying ML models.
However, since it was not possible to find any more recent prior works
evaluated on Arria 10 devices that also evaluate identical
ML models as we did, we also provide Table \ref{tab:mixed}, which compares prior
works sometimes evaluated on different FPGAs than our validation platform, but
for identical ML models and identical or similar input/datapath bitwidths.

Despite the strategies used in some prior works for increasing the theoretical
limits of the performance metrics defined in \equations{\ref{metric:thpt}} -
\eqPost{\ref{metric:thpt-area-cc}}, the results from Tables \ref{tab:first} -
\ref{tab:mixed}, demonstrate that the proposed FFIP approach is the most
effective at doing so and can provide performance improvements against the
best-in-class prior works in almost all evaluated performance metrics and every
evaluated comparison criteria. When comparing the lowest to highest performing
models evaluated on FFIP for \opsAreaCC against the
next-most competitive result from the prior works in each table,
FFIP is approximately \eOpsAreaCC
higher in Table \ref{tab:8}, it is overall on-par in Table \ref{tab:16} (even when including
the CPU-FPGA heterogeneous system in the comparison \cite{jiang2022cpu}), and
\mixedOpsAreaCC higher in Table \ref{tab:mixed}. Furthermore, once we account
for the higher clock frequency achieved due to the inherent hardware-efficient
nature of FFIP, it is further improved over the next-most
competitive works in each table, achieving approximately a \eOpsArea improvement
in Table \ref{tab:8}, a \sOpsArea improvement in Table \ref{tab:16}, and a
\mixedOpsArea improvement in Table \ref{tab:mixed}. Finally, our FFIP
implementation allowed us to achieve the highest overall throughput over the
next-most competitive prior works in each table, achieving approximately a \eOps
improvement in Table \ref{tab:8}, a \sOps improvement in Table \ref{tab:16}, and
a \mixedOps improvement in Table \ref{tab:mixed}, allowing us to increase the
throughput limits of the evaluation device well beyond its theoretical
throughput limits determined by \eq{\ref{thpt-roof-baseline}}.

%% file: results-8b-table.tex
%Please add the following packages if necessary:
%\usepackage{soul}% for underlines
%\usepackage[table]{xcolor} % for cell colors
%If the table is too wide, replace
%% \begin{table}[!htp]...\end{table} with

%\begin{adjustwidth}{-2.5 cm}{-2.5 cm}\centering\begin{threeparttable}[!htb]...\end{threeparttable}\end{adjustwidth}
%% \npdecimalsign{.}
%% \nprounddigits{2}

\begin{table*}[]\centering
%% \begin{adjustwidth}{-2.5 cm}{-2.5 cm}\centering\begin{threeparttable}[!htb]
%% \sisetup{round-mode=places}
%% \sisetup{table-column-width=16ex,    % local setting for all S columns
%%     round-mode=places,round-precision=3
%% }
\sisetup{round-precision=2}          % change setting for next S columns

%% NOTE: the width of each cell can be adjusted by modifying the following line in master:
%% \newcolumntype{B}{>{\centering\arraybackslash}p{1.5cm}}
%% NOTE: By changing a column type to 'H' below after tabular, it will hide that column
\caption{Comparison with state-of-the-art 8-bit-input accelerators for different models on the same FPGA family.}
\label{tab:8}
\label{tab:first}
\scriptsize
\begin{threeparttable}
\begin{tabular}{|>{\raggedleft}p{1.8cm}|DA|DA|DA||BDDD|}\toprule
\arrayrulecolor{black}
                                   &\multicolumn{2}{c|}{TNNLS '22 \cite{liu2022toward}} &\multicolumn{2}{c|}{TCAD '22 \cite{fan2022fpga}} &\multicolumn{2}{c||}{Entropy '22 \cite{an2022opencl}} &\multicolumn{4}{c|}{Ours (FFIP  64\by64)} \\
\toprule
FPGA                               &\multicolumn{2}{c|}{\gx}                            &\multicolumn{2}{c|}{\gx}                         &\multicolumn{2}{c||}{\gx}                             &\multicolumn{4}{c|}{\gx} \\
\arrayrulecolor{black!30}\midrule
\arrayrulecolor{black}
Data type                          &\multicolumn{2}{c|}{8-bit fixed}                    &\multicolumn{2}{c|}{8-bit fixed}                 &\multicolumn{2}{c||}{8-bit fixed}                     &\multicolumn{4}{c|}{8-bit fixed} \\
\arrayrulecolor{black!30}\midrule
\arrayrulecolor{black}
ALMs                               &\multicolumn{2}{c|}{304K}                           &\multicolumn{2}{c|}{304K}                        &\multicolumn{2}{c||}{303K}                            &\multicolumn{4}{c|}{\eALMs}        \\
\arrayrulecolor{black!30}\midrule
\arrayrulecolor{black}
Registers                          &\multicolumn{2}{c|}{889K}                           &\multicolumn{2}{c|}{890K}                        &\multicolumn{2}{c||}{-}                               &\multicolumn{4}{c|}{\eRegs}     \\
\arrayrulecolor{black!30}\midrule
\arrayrulecolor{black}
Memories                           &\multicolumn{2}{c|}{2334}                           &\multicolumn{2}{c|}{2334}                        &\multicolumn{2}{c||}{1953}                            &\multicolumn{4}{c|}{\eMems}        \\
\arrayrulecolor{black!30}\midrule
\arrayrulecolor{black}
DSPs                               &\multicolumn{2}{c|}{1473}                           &\multicolumn{2}{c|}{1473}                        &\multicolumn{2}{c||}{1503}                            &\multicolumn{4}{c|}{\eDSPs}        \\
\arrayrulecolor{black!30}\midrule
\arrayrulecolor{black}
Frequency (MHz)                    &\multicolumn{2}{c|}{200}                            &\multicolumn{2}{c|}{220}                         &\multicolumn{2}{c||}{172}                             &\multicolumn{4}{c|}{\eFreq} \\
\arrayrulecolor{black!30}\midrule
\arrayrulecolor{black}
Model                              &ResNet-50      &VGG16                               &Bayes ResNet-18  &Bayes VGG11                    &R-CNN (ResNet-50) &R-CNN (VGG16)                      &AlexNet             &ResNet-50            &ResNet-101          &ResNet-152 \\
\arrayrulecolor{black}\midrule
\thpt                              &1519           &1295                                &1590             &534                            &719               &865                                &\eAlexNetGOPS       &\eResNetAGOPS        &\eResNetBGOPS       &\eResNetCGOPS       \\
\arrayrulecolor{black!30}\midrule
\arrayrulecolor{black}
\gpmac                             &0.258          &0.220                               &0.270            &0.091                          &0.239             &0.288                              &\eAlexNetGOPSArea   &\eResNetAGOPSArea    &\eResNetBGOPSArea   &\eResNetCGOPSArea   \\
\arrayrulecolor{black!30}\midrule
\arrayrulecolor{black}
\macUt                             &1.289          &1.099                               &1.277            &0.412                          &1.391             &1.673                              &\eAlexNetGOPSAreaCC &\eResNetAGOPSAreaCC  &\eResNetBGOPSAreaCC &\eResNetCGOPSAreaCC \\
\arrayrulecolor{black}
\bottomrule
\end{tabular}
\begin{tablenotes}
  \thptExpl
\macPerDspExpl
\macUtExpl
\end{tablenotes}
\end{threeparttable}
\end{table*}

%% file: results-16b-table.tex
%Please add the following packages if necessary:
%\usepackage{soul}% for underlines
%\usepackage[table]{xcolor} % for cell colors
%If the table is too wide, replace
%% \begin{table}[!htp]...\end{table} with

%\begin{adjustwidth}{-2.5 cm}{-2.5 cm}\centering\begin{threeparttable}[!htb]...\end{threeparttable}\end{adjustwidth}
%% \npdecimalsign{.}
%% \nprounddigits{2}

\begin{table*}[]\centering
%% \begin{adjustwidth}{-2.5 cm}{-2.5 cm}\centering\begin{threeparttable}[!htb]
%% \sisetup{round-mode=places}
%% \sisetup{table-column-width=16ex,    % local setting for all S columns
%%     round-mode=places,round-precision=3
%% }
\sisetup{round-precision=2}          % change setting for next S columns

%% NOTE: the width of each cell can be adjusted by modifying the following line in master:
%% \newcolumntype{B}{>{\centering\arraybackslash}p{1.5cm}}
%% NOTE: By changing a column type to 'H' below after tabular, it will hide that column
\caption{Comparison with state-of-the-art 16b-bit-input accelerators for different models on the same FPGA family.}
\label{tab:16}
\scriptsize
\begin{threeparttable}
\begin{tabular}{|>{\raggedleft}p{1.8cm}|BDA|ZB|B|B||AAAA|}\toprule
\arrayrulecolor{black}
                                   &\multicolumn{3}{c|}{TCAD '20 \cite{ma2020automatic}}  &\multicolumn{2}{c|}{TVLSI '20 \cite{yepez2020stride}} &TCAS-II '22 \cite{jiang2022cpu} &TCAS-I '23 \cite{kim2023cnn} &\multicolumn{4}{c|}{Ours (FFIP  64\by64)} \\
\toprule
FPGA                               &\multicolumn{3}{c|}{\gx}                              &\multicolumn{2}{c|}{Arria 10}                        &\gx                             &Arria 10 SoC                 &\multicolumn{4}{c|}{\gx} \\
\arrayrulecolor{black!30}\midrule
\arrayrulecolor{black}
Data type                          &\multicolumn{3}{c|}{16-bit fixed}                     &\multicolumn{2}{c|}{16-bit fixed}                    &8/16-bit fixed \tnote{4}        &16-bit fixed                 &\multicolumn{4}{c|}{16-bit fixed} \\
\arrayrulecolor{black!30}\midrule
\arrayrulecolor{black}
ALMs                               &286K       &335K        &208K                          &181K   &180K                                        &-                               &189K                         &\multicolumn{4}{c|}{\sALMs}        \\
\arrayrulecolor{black!30}\midrule
\arrayrulecolor{black}
Registers                          &-          &-           &-                             &-      &-                                           &-                               &-                            &\multicolumn{4}{c|}{\sRegs}     \\
\arrayrulecolor{black!30}\midrule
\arrayrulecolor{black}
Memories                           &2356       &2692        &2319                          &1310   &1310                                        &1565                            &-                            &\multicolumn{4}{c|}{\sMems}        \\
\arrayrulecolor{black!30}\midrule
\arrayrulecolor{black}
DSPs                               &1518       &1518        &1518                          &1344   &1344                                        &1161                            &1536                         &\multicolumn{4}{c|}{\sDSPs}        \\
\arrayrulecolor{black!30}\midrule
\arrayrulecolor{black}
Frequency (MHz)                    &240        &240         &240                           &250    &250                                         &163                             &200                          &\multicolumn{4}{c|}{\sFreq} \\
\arrayrulecolor{black!30}\midrule
\arrayrulecolor{black}
Model                              &ResNet-50  &ResNet-152  &VGG16                         &VGG16  &Modified VGG16                              &CTPN( VGG +BiLSTM)              &Modified StyleNet            &AlexNet             &ResNet-50            &ResNet-101          &ResNet-152 \\
\arrayrulecolor{black}\midrule
\thpt                              &600        &697         &968                           &1642   &1788                                        &1224                            &670                          &\sAlexNetGOPS       &\sResNetAGOPS        &\sResNetBGOPS       &\sResNetCGOPS        \\
\arrayrulecolor{black!30}\midrule
\arrayrulecolor{black}
\gpmac                             &0.198      &0.230       &0.319                         &0.611  &0.655                                       &0.527                           &0.218                        &\sAlexNetGOPSArea   &\sResNetAGOPSArea    &\sResNetBGOPSArea   &\sResNetCGOPSArea    \\
\arrayrulecolor{black!30}\midrule
\arrayrulecolor{black}
\macUt                             &0.823      &0.957       &1.329                         &2.443 \tnote{5}&2.661 \tnote{5}                     &3.234 \tnote{5} \tnote{6}       &1.090                        &\sAlexNetGOPSAreaCC &\sResNetAGOPSAreaCC  &\sResNetBGOPSAreaCC &\sResNetCGOPSAreaCC  \\
\arrayrulecolor{black}
\bottomrule
\end{tabular}
  \begin{tablenotes}
\footNoteRefs
\item[4] Weights and layer outputs are quantized to 8 bits. Layer input is quantized to 16-bit due to Winograd convolutional transformations.
\item[5] These works use Winograd's minimal filtering algorithms \cite{lavin_fast} to reduce multiplication complexity.
\item[6] This is a central processing unit (CPU)-FPGA heterogeneous work where portions of the inference are performed on CPU.
\end{tablenotes}
\end{threeparttable}
\end{table*}

%% file: results-mixed.tex
\begin{table*}[]\centering
%% \begin{adjustwidth}{-2.5 cm}{-2.5 cm}\centering\begin{threeparttable}[!htb]
%% \sisetup{round-mode=places}
%% \sisetup{table-column-width=16ex,    % local setting for all S columns
%%     round-mode=places,round-precision=3
%% }
\sisetup{round-precision=2}          % change setting for next S columns

%% NOTE: the width of each cell can be adjusted by modifying the following line in master:
%% \newcolumntype{B}{>{\centering\arraybackslash}p{1.5cm}}
%% NOTE: By changing a column type to 'H' below after tabular, it will hide that column
\caption{Comparison with state-of-the-art accelerators on different FPGAs for the same models and input bitwidths.}
\label{tab:mixed}
\label{tab:last}
\scriptsize
\begin{threeparttable}
  \begin{tabular}{|>{\raggedleft}p{1.8cm}||A|A|A||A|A|A||A|A||A|A||A|A||}\toprule
    \arrayrulecolor{black}
                                    &TVLSI '19 \cite{kala2019high} &TCAS-II '21 \cite{li2021fpga} &Ours (FFIP  64\by64) &TNNLS '22 \cite{liu2022toward} &TCAS-I '23 \cite{kim2023agamotto} &Ours (FFIP  64\by64) &TCAD '20 \cite{ma2020automatic}&Ours (FFIP  64\by64) &TNNLS '22 \cite{huang2022fpga} &Ours (FFIP  64\by64) &TCAD '20 \cite{ma2020automatic}&Ours (FFIP  64\by64) \\
\toprule
Model                               &AlexNet                      &AlexNet                       &AlexNet              &ResNet-50                      &ResNet-50                         &ResNet-50            &ResNet-50                      &ResNet-50            &ResNet-101                     &ResNet-101           &ResNet-152                     &ResNet-152           \\
\arrayrulecolor{black!30}\midrule
\arrayrulecolor{black}
Data type                           &16-bit fixed                 &8/16-bit fixed \tnote{4}      &16-bit fixed         &8-bit fixed                    &8-bit fixed                       &8-bit fixed          &16-bit fixed                   &16-bit fixed         &8/16-bit fixed  \tnote{5}      &16-bit fixed         &16-bit fixed                   &16-bit fixed         \\
\arrayrulecolor{black!30}\midrule
\arrayrulecolor{black}
FPGA                                &XC7VX 690T                   &VC709                         &\gx                  &\gx                            &XCVU9P                            &\gx                  &\gx                            &\gx                  &VX980                          &\gx                  &\gx                            &\gx                  \\
\arrayrulecolor{black!30}\midrule
\arrayrulecolor{black}
ALMs (Intel) / LUTs (AMD)           &468K                         &121K                          &\sALMs               &304K                           &-                                 &\eALMs               &286K                           &\sALMs               &480K                           &\sALMs               &335K                           &\sALMs               \\
\arrayrulecolor{black!30}\midrule
\arrayrulecolor{black}
Registers                           &649K                         &160K                          &\sRegs               &889K                           &-                                 &\eRegs               &-                              &\sRegs               &-                              &\sRegs               &-                              &\sRegs               \\
\arrayrulecolor{black!30}\midrule
\arrayrulecolor{black}
Memories  (20Kb Intel) / (36Kb AMD) &1465                         &1470                          &\sMems               &2334                           &-                                 &\eMems               &2356                           &\sMems               &1457                           &\sMems               &2692                           &\sMems               \\
\arrayrulecolor{black!30}\midrule
\arrayrulecolor{black}
DSPs                                &1436                         &664                           &\sDSPs               &1473                           &2048                              &\eDSPs               &1518                           &\sDSPs               &3121                           &\sDSPs               &1518                           &\sDSPs               \\
\arrayrulecolor{black!30}\midrule
\arrayrulecolor{black}
Frequency (MHz)                     &200                          &200                           &\sFreq               &200                            &200                               &\eFreq               &240                            &\sFreq               &100                            &\sFreq               &240                            &\sFreq               \\
\arrayrulecolor{black}\midrule
\thpt                               &434                          &220                           &\sAlexNetGOPS        &1519                           &287                               &\eResNetAGOPS        &600                            &\sResNetAGOPS        &600                            &\sResNetBGOPS        &697                            &\sResNetCGOPS        \\
\arrayrulecolor{black!30}\midrule
\arrayrulecolor{black}
\gpmac                              &0.302                        &0.331                         &\sAlexNetGOPSArea    &0.258                          &0.140                             &\eResNetAGOPSArea    &0.198                          &\sResNetAGOPSArea    &0.192                          &\sResNetBGOPSArea    &0.230                          &\sResNetCGOPSArea    \\
\arrayrulecolor{black!30}\midrule
\arrayrulecolor{black}
\macUt                              &1.511                        &1.657                         &\sAlexNetGOPSAreaCC  &1.289                          &0.701                             &\eResNetAGOPSAreaCC  &0.823                          &\sResNetAGOPSAreaCC  &1.922                          &\sResNetBGOPSAreaCC  &0.957                          &\sResNetCGOPSAreaCC  \\
\arrayrulecolor{black}
\bottomrule
\end{tabular}
\begin{tablenotes}
\footNoteRefs
\item[4] Weights and layer inputs are quantized to 8 and 16 bits, respectively.
\item[5] Weights are quantized to 8 bits and layer input/output is quantized to 8 or 16 bits at different stages.
\end{tablenotes}
\end{threeparttable}
\end{table*}

%% file: conclusion.tex
\section{Conclusion}
We present an algorithm and general architecture that improve Winograd's
under-explored inner-product algorithm \cite{wino} that can be seamlessly
incorporated into any ML accelerator system that uses traditional fixed-point
systolic arrays to double the throughput per MAC unit,
significantly increasing the accelerator's performance per compute area across
all ML models that will execute on the systolic array.

We implement and evaluate FIP for the first time in a machine learning
accelerator system. We then identify a weakness of FIP and propose
the new FFIP algorithm and generalized hardware architecture that inherently
address that weakness in the general case. We provide ML-specific optimizations
for the FIP and FFIP algorithms and systolic array hardware architectures. We
derive how the (F)FIP architectures increase the theoretical compute efficiency
and performance limits in the general case.

Finally, although the theoretical concepts presented in this work are general
and applicable to both custom integrated circuits and FPGA implementations,
since our proposed FFIP algorithm and architecture were validated on FPGA,
we empirically confirm
our theoretical insights through comparison with prior state-of-the-art
solutions also evaluated on FPGA. As anticipated from our theoretical analysis,
our full-system validation results shown in \Figure \ref{fig:report} and Tables
\ref{tab:first} - \ref{tab:last} confirm that the generalized FFIP systolic
array architecture we propose, when implemented, does in-fact surpass the
traditional theoretical compute efficiency and performance limits, and can
improve performance compared to the best-in-class prior
state-of-the-art solutions in almost all evaluated performance
metrics, each evaluated comparison criteria, and all evaluated ML models.
More importantly, our results indicate that FFIP
can directly benefit the ML acceleration community by allowing for
an increase in the theoretical compute efficiency and performance limitations in
the general case across a wide range of devices, system implementations, and ML
models.